\documentclass[]{aa}
\usepackage{txfonts}
\usepackage{graphicx}
\usepackage{epsfig}
\usepackage{rotating}
\usepackage{amssymb}
\usepackage{natbib}
\bibpunct{(}{)}{;}{a}{}{,}
\usepackage{txfonts}
\begin{document}
\title{Diagnostics of non-thermal distributions in solar flare spectra observed by
RESIK and RHESSI\thanks{Appendix A is available at electronic form at http://www.aanda.org}}
\author{A. Kulinov\'{a}\inst{\ref{inst1},\ref{inst2}}
\and J. Ka\v{s}parov\'{a}\inst{\ref{inst1}}
\and E. Dzif\v{c}\'{a}kov\'{a}\inst{\ref{inst1}}
\and J.~Sylwester\inst{\ref{inst3}}
\and B. Sylwester\inst{\ref{inst3}} 
\and M. Karlick\'y\inst{\ref{inst1}}}

\institute{Astronomical Institute of the Academy of Sciences of the Czech Republic, v.~v.~i.,
Fri\v{c}ova 298, 251 65 Ond\v{r}ejov, Czech Republic 
\email{[kulinova; kasparov; elena; karlicky]@asu.cas.cz}\label{inst1}
\and
Department of Astronomy, Physics of the Earth and Meteorology,
Faculty of Mathematics, Physics and Informatics, \\
Comenius University, Mlynsk\'{a} dolina, 842 48 Bratislava, Slovakia\\
\email{kulinova@fmph.uniba.sk}\label{inst2}
   \and
   Space Research Centre, Polish Academy of Sciences,
           51-622, Kopernika 11, Wroc{\l}aw, Poland\\
  \email{[js; bs]@cbk.pan.wroc.pl}\label{inst3}}

\date{Received  17 February 2011 / Accepted 7 July 2011}

\abstract
{During 
solar flares an enormous amount of energy is released, and the charged particles, like electrons, are accelerated. These non-thermal electrons interact with the plasma in various parts of solar flares, where the distribution function of electrons can therefore be non-Maxwellian.}
{We focus on the non-thermal components of the electron distribution in the keV range and analyse high-energy resolution X-ray spectra detected by RESIK and RHESSI for three solar flares. }
{In the 2\,--\,4~keV range we assume that the electron distribution can be modelled by an n-distribution. Using a method of line-intensity ratios, we analyse allowed and satellite lines of Si observed by RESIK and estimate the parameters of this n-distribution. At higher energies we explore RHESSI bremsstrahlung spectra. Adopting a forward-fitting approach and thick-target approximation, we determine the characteristics of injected electron beams.}
{RHESSI non-thermal component associated with the electron beam is correlated well with presence of the non-thermal n-distribution obtained from the RESIK spectra. In addition, such an n-distribution occurs during radio bursts 
observed in the 0.61\,--\,15.4~GHz range. Furthermore, we show that the n-distribution could also explain RHESSI emission below $\sim$\,5~keV. Therefore, two independent diagnostics methods indicate the flare plasma being affected by the electron beam can have a non-thermal component in the $\sim$\,2\,--\,5~keV range, which is described by the n-distribution well. Finally, spectral line analysis reveals that the n-distribution does not occupy the same location as the thermal component detected by RHESSI at $\sim$\,10~keV.}{}
\titlerunning{Diagnostics of non-thermal distributions in solar flares}

\authorrunning{Kulinov\'{a} et al.}
\maketitle
\newpage

\section{Introduction}

The energy release during the solar flares is accompanied by many phenomena, i.e. magnetic field reconnection and reconfiguration, particle acceleration, plasma heating, and electromagnetic radiation across the whole spectral range.

X-ray spectra in deka keV range allow us to diagnose a thermal part of the flare plasma, as well as its non-thermal, high-energy tail. Recent high-energy resolution, hard X-ray spectra obtained by Reuven Ramaty High-Energy Solar  Spectroscopic Imager,  RHESSI, \citep{lin02} enable us to infer the corresponding electron spectrum. This can be achieved with a model-independent approach, such as regularised techniques \citep{pia03,kont04}. Also, the forward-fitting method is often used after assuming that the spectrum of the electrons emitting in this energy range is the sum of an isothermal component and a non-thermal power-law distribution: e.g., \citet{brown71} and \citet{hol03}. This approach is adopted here.

In space plasma,  the non-thermal electron distributions are commonly modelled by $\kappa$-distributions, based on in situ measurements, e.g., \citet{van09,mak97}. \citet{tsa88} and \citet{leub02} have shown that the $\kappa$-distribution is a consequence of the generalised entropy favoured by non-extensive statistics, therefore having a deeper physical meaning. \citet{kasp09} interpreted RHESSI X-ray spectra of several flares using $\kappa$-distributions. They found that some loop-top X-ray sources can be described by such a distribution.
\begin{table*}
\centering
    \caption{Spectral lines used for diagnostics of the non-thermal distribution from RESIK spectrometer.}
    \renewcommand{\arraystretch}{1.2}
    \begin{tabular}[t]{ccccc}
      \hline \hline
      Line number & Ion & Wavelength [\AA] & Transition & Ratio\\
      \hline
      1 & \ion{Si}{xiv}  & 5.22 & $1s~^{2}S_{1/2}$\,--\,$3p~^{2}P_{3/2,1/2}$ & \ion{Si}{xiv} / \ion{Si}{xiii}\\
      2 & \ion{Si}{xiii} & 5.68 & $1s^{2}~^{1}S_{0}$\,--\,$1s3p~^{1}P_{1}$ & \ion{Si}{xiii} / \ion{Si}{xii}d\\
      3 & \ion{Si}{xii}d\tablefootmark{*} & 5.82 & $1s^{2}2p~^{2}P_{1/2,3/2}$\,--\,$1s2p(3P)~3p~^{2}D_{3/2,5/2}$  &\,--\,\\
      \hline \\
      \end{tabular}
    \label{tab4}
\tablefoot{
\tablefoottext{*}{The satellite line excited by di-electronic recombination.}
}
\end{table*}

\citet{sel87} analysed SOLFLEX line spectra of \ion{Fe}{xxv} and \ion{Fe}{xxiv} during several flares. They found that enhanced intensities of the satellite lines can be explained by the presence of non-thermal energy distributions that have higher and narrower peaks than the Maxwell distribution. This type of distribution functions has been used in the analysis of non-thermal distribution in laboratory plasma \citep{har79} as well. In this paper we  refer to them as n-distributions. Recently, \citet{dzifkar08} have introduced a simple model of a mono-energetic electron beam penetrating the solar atmosphere and creating a return current. They showed that adding a drift velocity corresponding to the electric current results in a non-thermal electron distribution that can be approximated by the n-distribution. Both the electron beam and the n-distribution influence the line intensities but each in a different way. The latter enhances the intensities of satellite lines, while the electron beam affects the intensities of allowed lines dominantly. \citet{dzi08} analysed spectra of a flare using this kind of non-thermal distribution. They showed that the spectra, calculated using the n-distribution with the parameters estimated from observations, could mimic the observed relative intensities of allowed and satellite lines very well.
Additionally, they showed that the high intensities of the satellite lines observed
by RESIK cannot be reproduced by a multi-thermal distribution, which further supports the preferred choice of the n-distribution.

In this paper we apply two independent diagnostic techniques to analyse soft and hard X-ray flare spectra. In the soft X-ray range ($\sim$\,2~keV), where the bulk of the electron distribution is important, we consider the n-distribution. In the hard X-ray range (above 6 keV), we assume a single power-law distribution describing the high-energy tail. A Maxwell distribution 
accounts for a possible thermal component at keV energies. 
We analyse flare spectra detected by two spectrometers and determine the parameters of these model distribution functions.

\section{Model distribution functions}
In the present work we consider three types of distributions: a Maxwell, a single power law,  and an n-distribution, and for their determination we use different diagnostic methods. The thermal component is generally present and dominates the X-ray emission around 10~keV. We suppose it is an isothermal emission so we approximate the corresponding electron distribution function as Maxwell.

We assume that the hard X-ray emission is caused by the thick-target bremsstrahlung \citep{brown71} of an electron beam injected into the plasma with a single power-law spectrum:
\begin{equation}\label{eq:power-law}
 F(E)=(\delta -1)\frac{F_\mathrm{T}}{E_\mathrm{C}} \left( \frac{E}{E_{\mathrm{C}}} \right )^{-\delta} = C.E^{-\delta},
\end{equation}
where $\delta$ is the spectral index, $E_{\mathrm{C}}$ the low-energy cutoff,
$F_\mathrm{T}$ represents the 
total rate of injected electrons of $E \ge E_{\mathrm{C}}$, and $C$ is a constant. Symbol $F(E)$ in Eq.~\ref{eq:power-law} 
is the electron rate spectrum \citep{brown05} in units [s$^{-1}$ eV$^{-1}$].

\begin{figure}
 \centering \includegraphics[width=8.0cm]{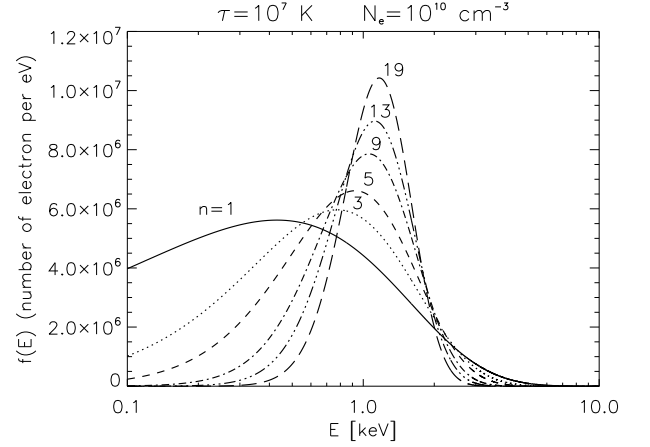}
\caption{n-distribution functions for different values of the parameter $n$=1, 3, 5, 9, 13, 19. 
$n=1$ represents the Maxwell distribution.}
\label{fig_ndistr}
\end{figure}

At the lower energies of X-ray flare 
spectra, i.e. 2\,--\,2.5 keV, we use a diagnostics based on soft X-ray line spectra without continuum. We follow the diagnostic technique developed by \citet{dzi08}, which uses the normalized n-distribution (Fig.~\ref{fig_ndistr}) to describe the electron distribution: 
\begin{equation}
f_\mathrm{n}(E) dE=B^\mathrm{n} \frac {2}{\sqrt {\pi}} {E}^{\frac{n}{2}}
(kT)^{-(\frac{n}{2}+1) }{\rm exp}(-E/kT) dE,
\end{equation}
where $B^\mathrm{n}= \pi^{1/2} /(2 \Gamma (\frac{n}{2}+1))$ is the normalization constant, $E$ the energy of free electrons, $k$ Boltzmann constant, and $n$ and $T$ are the free parameters of the distribution. The parameter $T$ in this distribution is not the thermodynamic temperature, although it is also given in Kelvin. The parameter $n$ is a dimensionless number describes the degree of deviation from the Maxwell distribution function. If $n$\,=1, the distribution function becomes Maxwell. The mean energy of the n-distribution, Eq.~\ref{eq:En}, depends on two parameters $n$ and $T$. To compare 
the non-thermal component with the 
thermal one, we use
the pseudo-temperature, $\tau$, which relates the mean energy of the n-distribution to the mean energy of the Maxwell distribution \citep{dzi98, dziku01}:
\begin{equation}
E_\mathrm{n}~=~(n+2)kT/2~=~3k\tau/2,
\label{eq:En}
\end{equation}
where $\tau=(n+2)T/3$ .

\subsection{Diagnostics of the n-distribution}\label{lineratiomethod}

\begin{figure}
 \centering \includegraphics[width=8.0cm]{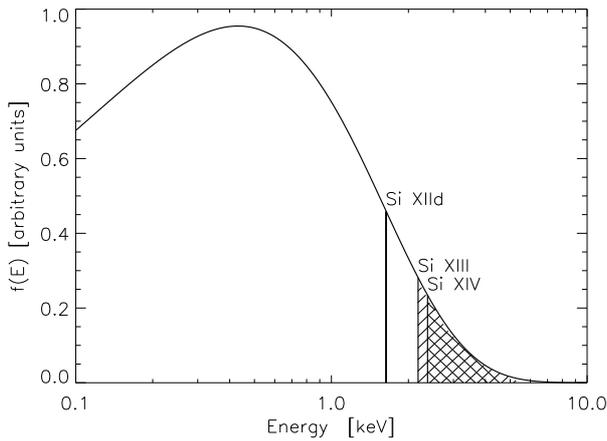}
\caption{Schematic figure demonstrates which parts of a distribution function influence the intensities of spectral lines (Table~\ref{tab4}) used for diagnostics of the n-distribution. The hatched regions of the distribution influence the intensities of the allowed lines of ions \ion{Si}{xiv} and \ion{Si}{xiii}. The discrete energy of the doubly excited state of \ion{Si}{xii}d is marked by a thick vertical line.}
\label{fig_sch}
\end{figure}

The line ratios of the three spectral lines listed in Table~\ref{tab4} were used for the diagnostics of the n-distribution.The intensities of two allowed lines (\ion{Si}{xiv} and \ion{Si}{xiii}) depend on excitation rates that are integrals 
of a collisional cross section, velocity, and an electron distribution function.
The intensities of the allowed lines are thus sensitive to the high-energy tail part of the electron distribution function (Fig.~\ref{fig_sch}). 

\begin{figure}
\centering \includegraphics[width=8.0cm]{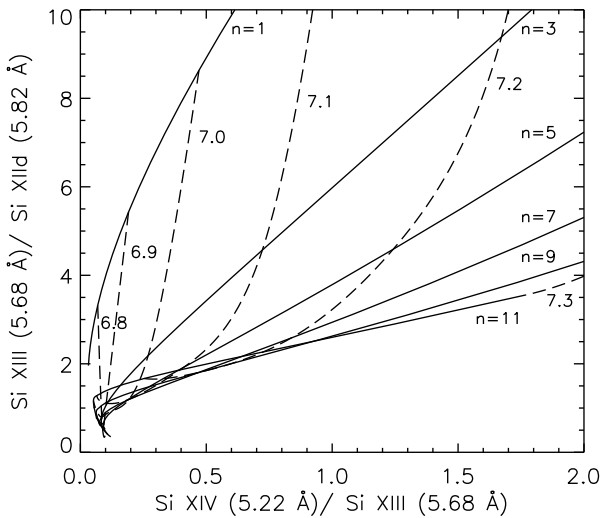}
\caption{Synthetic line ratios are functions of both $n$ and $\tau$. The plot shows
the synthetic line ratios for parameter $n$\,=\,1, 3, 5, 7, 9, 11, and the curves of constant log~$\tau$ (grey numbers). 
The curve $n$\,=\,1 represents the line ratios for the Maxwell distribution.}
\label{fig_syntrat}
\end{figure}

Besides the excitation equilibrium, the intensity ratio of the allowed lines is affected by the ionization equilibrium as well. The electron ionization rate of the \ion{Si}{xiii} is also an integral quantity that depends on the shape of the electron distribution function above the ionization energy of \ion{Si}{xiii}. However, the radiative recombination rate depends on the shape of the whole electron distribution function. Consequently, the ratio of the allowed lines is mainly sensitive to the pseudo-temperature $\tau$ but only weakly depends on the parameter $n$, see Fig.~\ref{fig_syntrat}. 
The \ion{Si}{xiv} / \ion{Si}{xiii} ratio is very sensitive to $T$ or $\tau$, and it varies from $\sim0.1$ for log $\tau$=6.8 to $\sim 2$
for log $\tau$=7.2. The magnitude of changes in this range of $\tau$ is smaller for the higher values of parameter n.
This behaviour corresponds to the changes in the ionization and excitation equilibrium because of the n-distribution
\citep{dzi98,dzito07}.
The satellite lines of \ion{Si}{xii} (further as \ion{Si}{xii}d) show different behaviours. Their intensities depend on atomic parameters and the number of electrons with the energy corresponding to the energy of the doubly excited state of ion \ion{Si}{xii} above its ionization limit (Fig.~\ref{fig_sch}). Thus, the ratio \ion{Si}{xiii}/\ion{Si}{xii}d is extremely sensitive to the shape of the electron distribution function. 
For detailed information see \citet{dzi11a}, who investigate the influence of the non-thermal plasma bulk and the high-energy tail  on the
Si ionization and its line spectra.
The observed enhanced intensities of satellite lines \citep{dzi08} require an  
increased number of particles at 
\ion{Si}{xii}d energy in comparison to the Maxwell distribution (Figs.~\ref{fig_ndistr}, \ref{fig_sch}, and \ref{fig_syntrat}). The n-distribution with higher $n$ satisfies this condition. 

Although the electron distribution is non-thermal, this does not necessarily imply that the ionization is out of equilibrium.
The ionization rates of Si XIV are of the order of $10^{-10}\,\mbox{cm}^3\mbox{s}^{-1}$, which results in the ionization time
of about 1~s for the electron density $10^{10}\,\mbox{cm}^{-3}$. As the results show 
\citep[this paper and][]{dzi08},
the typical time scale of changes in the derived plasma parameters is longer, several tens of seconds. 
Thus, the ionization equilibrium can be achieved if the conditions for formation of the non-thermal distributions 
change on a longer time scale than the ionization time. That seems to be the case for the present analysis, therefore the assumption
of the ionization equilibrium is valid here.

In our model the n-distribution represents the bulk of the plasma electron distribution. It does not include the enhanced high-energy tail, representing a beam of accelerated electrons, usually observed at hard X-ray energies.

We would like to point that, although the $\kappa$-distribution also provides the enhanced number of particles in its high-energy tail, its shape in the region of 1\,--\,4 keV is not able to explain the enhanced intensities of the satellite lines. This can occur only when the gradient of the electron distribution function is high enough in this energy region, and this is not fulfilled 
either by $\kappa$-distribution or the Maxwell one.

\section{Instruments and data analysis}

We 
analysed flare spectra observed by two X-ray spectrometers: REntgenovsky Spektrometr s Izognutymi Kristalami\,--\,RESIK \citep{syl05} and RHESSI.

\subsection{RESIK}

The first spectrometer, RESIK, is an uncollimated spectrometer with two bent quartz crystals that covers the 2.0\,--\,3.7~keV energy range. It has four spectral channels with nominal wavelength ranges: 3.4\,--\,3.8 \AA, 3.83\,--\,4.27 \AA, 4.35\,--\,4.86 \AA, and 5.00\,--\,6.05~\AA. 
In the present work we analysed spectra within the 5.15\,--\,5.95\,\AA~spectral range coming from the fourth channel where the allowed lines of ions \ion{S}{xv}, \ion{Si}{xiv}, \ion{Si}{xiii}, and the satellite lines of ion \ion{Si}{xii} dominate. The raw RESIK spectra were reduced to the absolute flux units by incorporating all the known instrumental factors \citep{syl05}. Then, the reduced spectra were time-weighted according to the exposition times of individual spectra. 
To improve the signal-to-noise ratio, we averaged the spectra in nearly equidistant time intervals, see Tables~\ref{tab1},~\ref{tab2}, and~\ref{tab3}. The diagnostic method of RESIK data is based on ratios of line fluxes, see Sect.~\ref{lineratiomethod}, therefore we subtracted a linear continuum from the time-weighted spectra. However, RESIK spectra are contaminated by an artificial continuum flux that are like mounds. The exact origin of these `mounds' is still unclear, but we removed them using
the same method as described in \citet{dzi08}. The imperfectly removed `mounds' can lead to relatively high errors of \ion{Si}{xiv} line intensity, in particular, thus causing higher uncertainty of $n$ for the spectra registered during a decay phase when this line intensity is low. Finally, the line spectra were fitted by Gaussian profiles with a constant full width at half maximum (FWHM). The errors of the line fluxes from the count statistics and fitting of the spectral lines reach 35\%. 

Diagnostics of the n-distribution from the RESIK spectra makes use of the precalculated grid of synthetic Si line ratios 
\citep{dzi08}. These synthetic line ratios are functions of both parameters $n$ and $\tau$, see Fig.~\ref{fig_syntrat}. Applying the spline interpolation to the synthetic line ratios, we determined the values of $n$ and log~$\tau$ with their errors from the measured line ratios. 

Our line ratio diagnostic method can reliably detect only $n$\,$\le$\,11. For higher values of $n$, the curves of constant $n$ start to overlap and lead to ambiguity in $n$. Therefore for all the measured line ratios lying below the curve $n$\,=\,11 (Fig.~\ref{fig_syntrat}), the parameter $n$ was set artificially to 11 and the upper limit of $n$ to 12. Such an upper limit was chosen only for display purposes, and physically it means that $n$\,$\ge$\,11. In contrast to the previous paper by \citet{dzi08}, log~$\tau$ was determined using both line ratios (Table~\ref{tab4}). 
There is a lower limit for $\log\tau$: Fig.~\ref{fig_syntrat} shows that only $\log\tau\ga 6.8$ can be reliably detected.
In the case of $n$\,=\,11, log~$\tau$ and its upper limit were determined as the values corresponding to $n$\,=\,11 and 12, respectively.

Obviously, the n-distribution with any parameter $n$\,>\,1 means a deviation from the Maxwell distribution. However, the errors from the count statistics and from removing the `mounds' mainly affect the intensities of weak lines, and the higher uncertainties can usually lead to overestimated values of $n$ during the decay phase. Therefore, in this study we concentrate on the time intervals of $n$\,$\ge$\,5.

\subsection{RHESSI}

The second spectrometer, RHESSI, was designed to investigate the flare spectra in the energy range from 3~keV up to 7~MeV. We were interested in non-thermal bremsstrahlung radiation produced by an electron beam injected to ambient plasma.
In all the analysed flares the enhanced hard X-ray emission was observed approximately up to 100 keV. Among the RHESSI flare spectra, we 
studied in detail those time intervals during which we could observe both thermal and non-thermal components. 

In the hard X-rays, the non-thermal components of the flares lasted from two to about four minutes. At later times the hard X-ray emission of the flares could be fitted by one or two thermal components. We assumed the emission is composed of continuum emission caused by electron-proton bremsstrahlung 
plus contributions of Fe ($\sim7$~keV) and Fe/Ni ($\sim8$~keV) line complexes \citep{phi04}.

We fitted the spectra in the Object SPectral EXecutive (OSPEX) environment using the model for the isothermal emission and lines below about 10~keV 
\citep[{\it vth} function with the default set of the solar coronal abundances,][]{fel92} and a thick-target non-thermal emission at higher energies 
\citep[{\it photon\_thick} function,][]{brownetal08}  assuming a single power-law electron beam, Eq.~(\ref{eq:power-law}). If necessary, the {\it pileup\_mod} pseudo-function\footnote{{http://prg.ssl.berkeley.edu/\~{}tohban/wiki/index.php/Pileup\_mod\_-\_Pseudo\_function\_for\_correcting\_pileup}} with fixed default parameters was used. For spectra without the attenuator, state A0, the {\it drm\_mod} pseudo-function\footnote{{http://sprg.ssl.berkeley.edu/\~{}tohban/wiki/index.php/Drm\_mod\_-\_Pseudo\_function\_for\_fine\_tuning\_RHESSI\_DRM\_parameters}} was applied. We analysed the output from a single RHESSI detector 4F only because of its fine energy resolution of 0.98~keV \citep{smith02}. 
Also, it produces a photon spectrum close to the mean spectrum of all front detectors usable for spectroscopy \citep{phi06}. Moreover, the {\it drm\_mod} function can only be used to fit single-detector spectra. The low-energy boundary for the fits was set according to the attenuator state: $\sim$ 6 keV in A0, $\sim 8$ keV in A1, and $\sim$ 12~keV in A3. The upper boundary of the energy fitting ranges was set to the values where the flare data were nearly equal to the background. In this way we obtained the time evolution of temperature of the thermal component $T$ and the parameters of the injected electron beam, $\delta$, $E_\mathrm{C}$, and $F_\mathrm{T}$.

\subsection{Radio data}

As complimentary data we used radio observations provided by USAF Radio Solar Telescope Network
(RSTN) and Ond\v{r}ejov radiospectrographs 
\citep{jir93}.
RSTN network consists of four solar radio observatories that monitor radio emission from the Sun on eight discrete frequencies (in GHz): 0.245, 0.410, 0.610, 1.415, 2.695, 4.995, 8.800, and 15.400. The solar radio flux at these frequencies is recorded each second and is given in SFU units.
Two radiospectrographs RT4 (2.0-4.5 GHz) and RT5 (0.8-2.0 GHz) at Ond\v{r}ejov, Czech Republic, observed the Sun with 
0.1~s time resolution and sample the frequency band in 256 frequency channels.

\section{Flare events}
\begin{figure*}
\centering {\includegraphics[width=8.00cm]{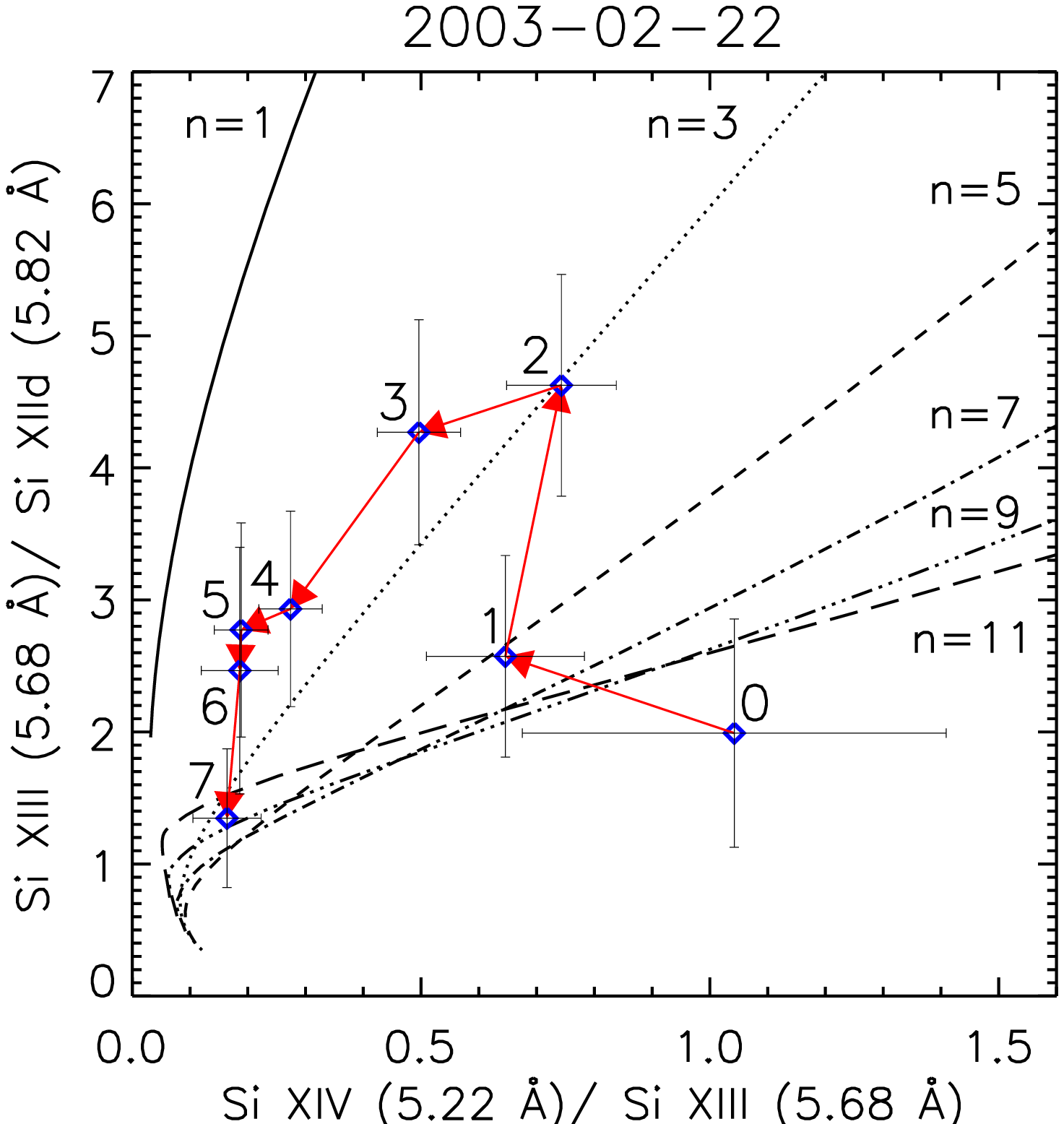} \includegraphics[width=8.00cm]{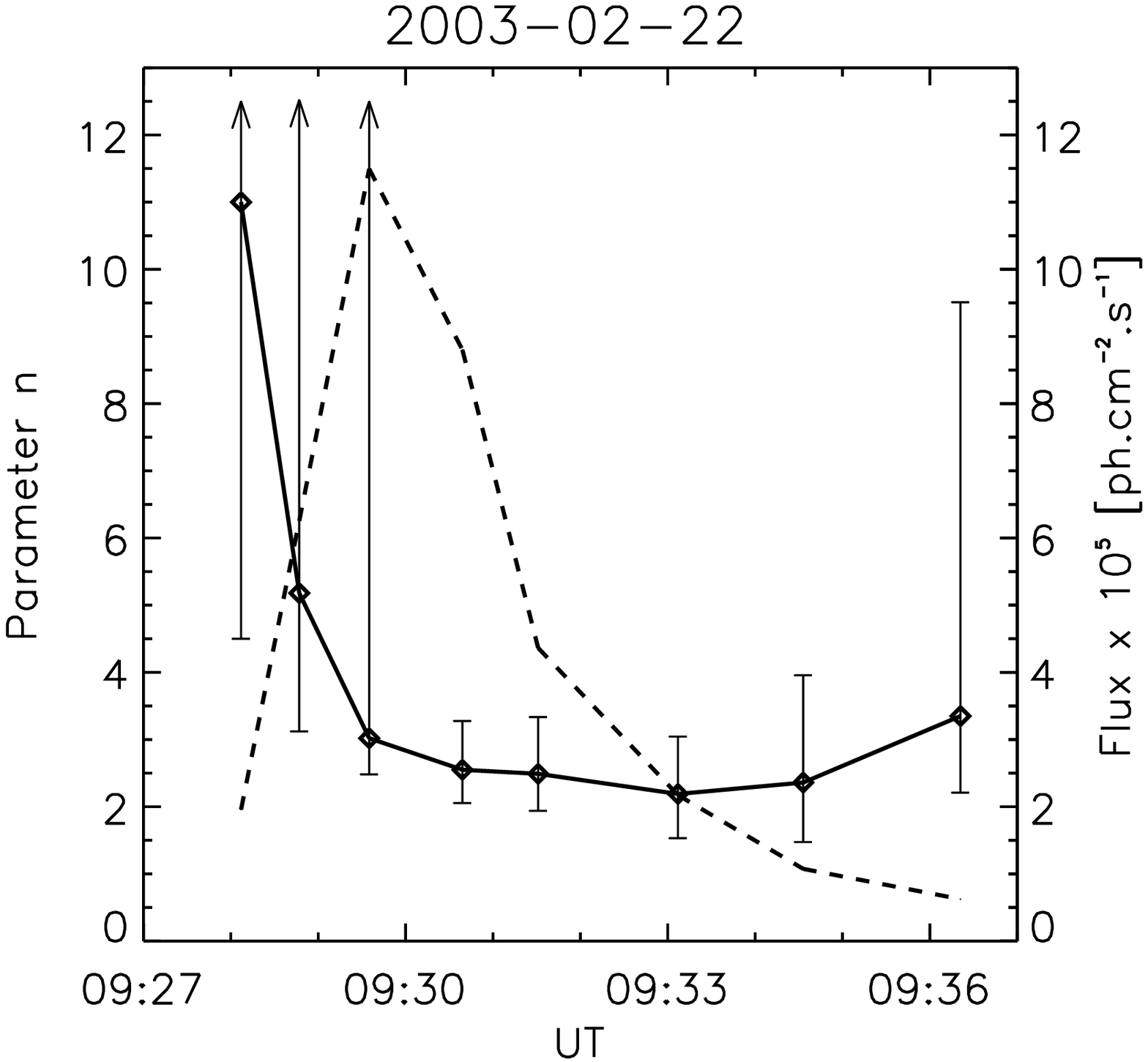}}
\centering{\includegraphics[width=8.00cm]{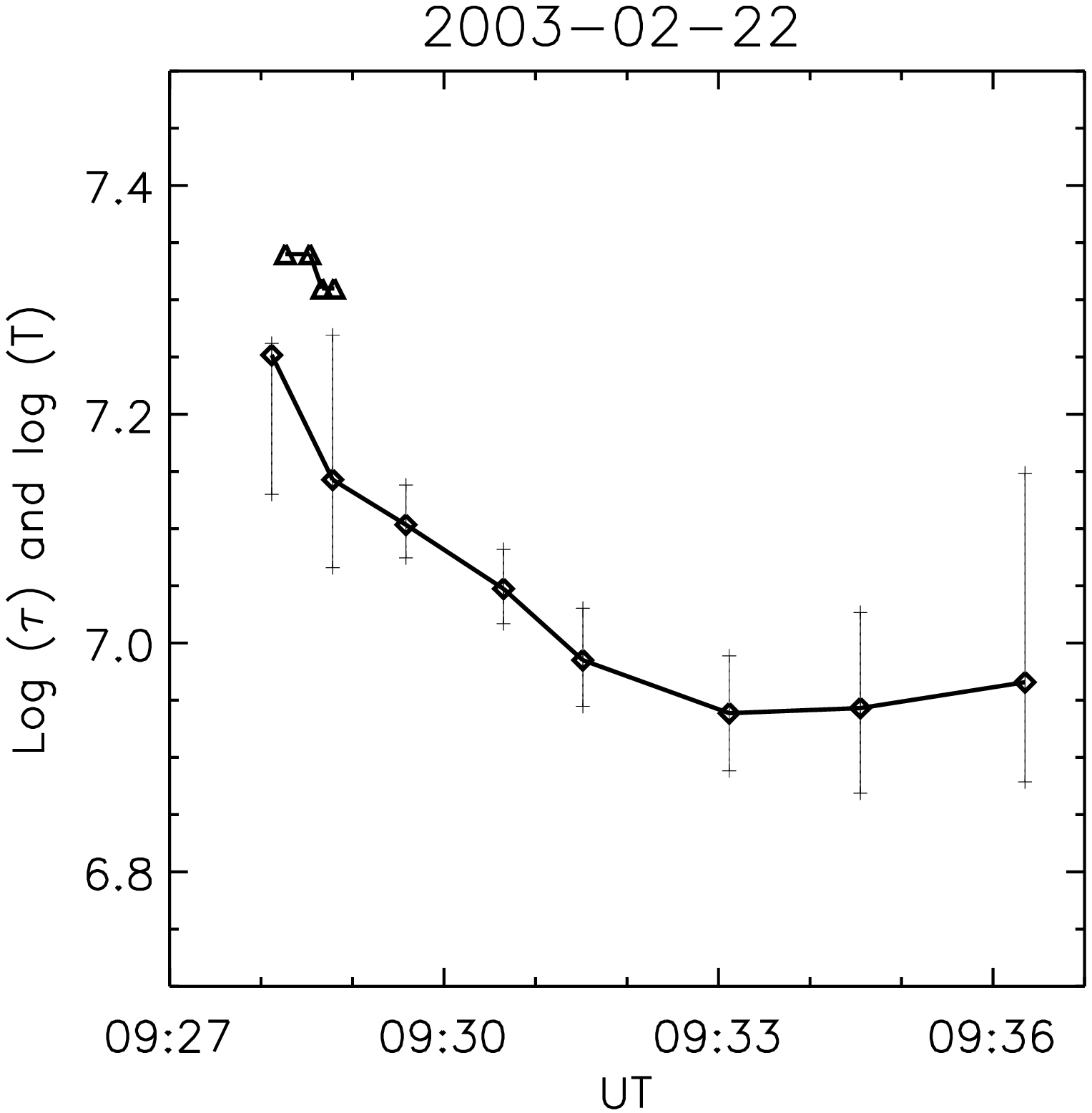}
\includegraphics[width=8.00cm]{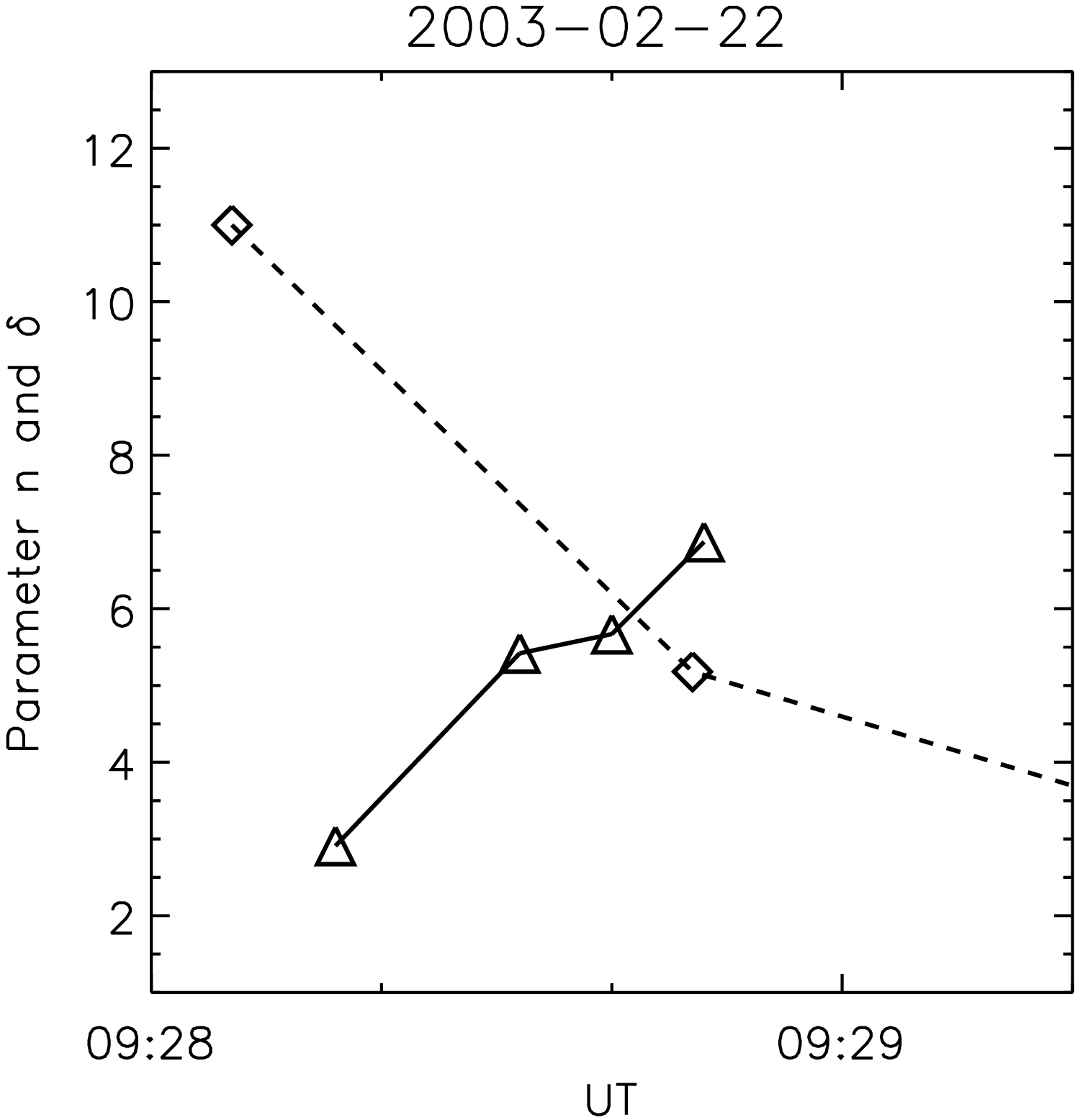}}
\caption{RESIK and RHESSI analysis results of the 2003 February 22 flare. Top left: the synthetic line ratios for $n$=1, 3, 5, 7, 9, 11 (solid, dotted, dashed, dot-dashed, dot-dot-dot dashed, and long-dashed lines, respectively), and the measured line ratios (stars) with their error bars. The arrows follow the time evolution of the RESIK spectra 
(0 marks the first observed spectrum, see Table~\ref{tab1}). Top right: the time evolution of 
the parameter $n$ (with error bars, upward arrows indicate $n>12$) of the n-distribution and the total flux (dashed line) from all four RESIK channels. Bottom left: the time evolution of log~$\tau$ (diamonds) from RESIK and log~$T$ (triangles) from the RHESSI thermal component. Bottom right: the time evolution of 
the spectral index $\delta$ (triangles) in comparison with the parameter $n$ (diamonds and dashed line). 
This panel zooms in on the time evolution of $n$ depicted in the panel above.
The points in all the plots are depicted for the middle time of each spectrum.}
\label{fig_res1}
\end{figure*}

RESIK detected several flares between August 2001 and April 2003. From its catalogue of flares 
we picked up three flares that were simultaneously observed by RHESSI and RSTN. One of the flares was observed by radiospectrographs at Ond\v{r}ejov Observatory.

\subsection{2003 February 22 flare}\label{sec_feb22}

The first flare was observed on February 22, 2003 in the active region NOAA 10290. According to Solar Event Reports
it was of GOES class C5.8 with the soft X-ray beginning at 09:24 UT, and with its maximum at 09:29 UT and end at 09:31 UT.

The results from the RESIK line spectra and RHESSI analysis are shown in Fig.~\ref{fig_res1}. The top left-hand panel shows the comparison of the \ion{Si}{xiv} / \ion{Si}{xiii} vs. \ion{Si}{xiii} / \ion{Si}{xii}d synthetic line ratios with the measured ones. 
The significant deviation of electron distribution from the thermal one occurred slightly before the GOES soft X-ray maximum and during the time of increasing RESIK X-ray flux (09:27:43\,--\,09:29:03 UT), see Fig.~\ref{fig_res1} (right column). Later on, 
$n\sim$~3, which suggests slow thermalization of the flare plasma.

Figure~\ref{fig_res1} (bottom left) shows that the log~$\tau$ decreased monotonically from its highest value, log~$\tau$\,$\sim$\,7.3 (1.5~keV) observed in the rise phase of the flare. The log~$T$ from RHESSI is higher by $\sim$\,0.1 dex than log~$\tau$. 
Tables~\ref{tab1} and \ref{tabx1} list the determined RESIK and RHESSI parameters, respectively.

\begin{figure}
\centering \includegraphics[width=9.0cm]{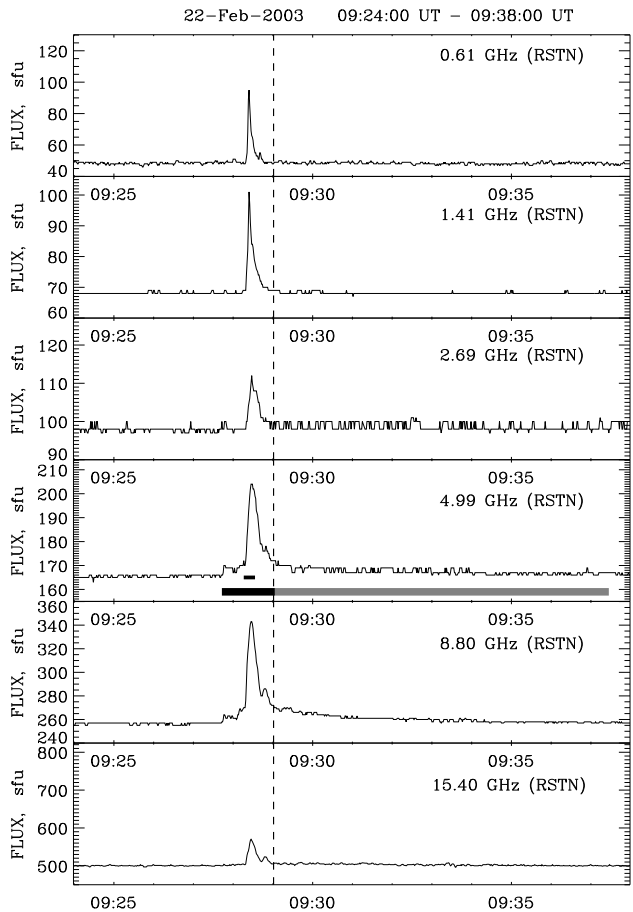}
\caption{Radio bursts observed on February 22, 2003 at six frequencies 15.40, 8.80, 4.99, 2.69, 1.41, and 0.61 GHz by RSTN. The dashed vertical line at 09:29 UT marks the time of the GOES soft X-ray flare maximum. The heavy black line denotes the time interval with the parameter $n\,\ge\,$5 determined from RESIK  and the heavy grey line represents time intervals with $n\,<\,5$. The short black line denotes the time interval when groups of dm-type III bursts were observed, see Fig.~\ref{fig_ond}.}
\label{fig_rstn1}
\end{figure}

\begin{figure}
\centering \includegraphics[width=8.8cm]{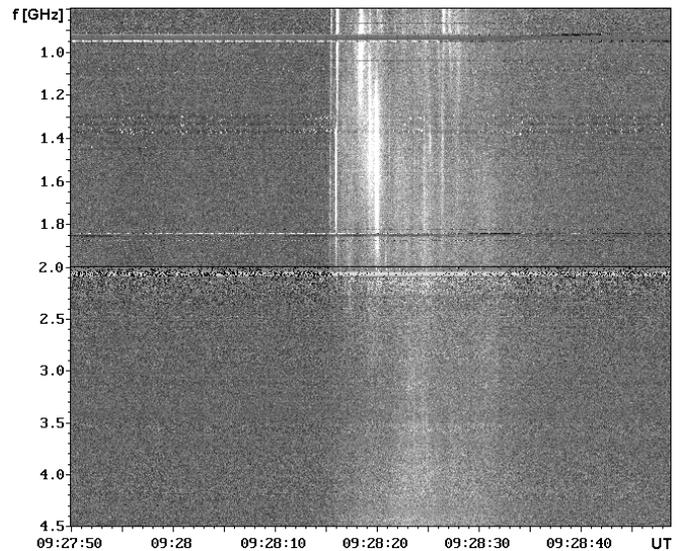}
\caption{Composite radio spectrum at 0.8\,--\,4.5 GHz observed on February 22, 2003 by the Ond\v{r}ejov radiospectrographs showing a group of type III bursts in the time interval 09:28:16\,--\,09:28:33~UT.}
\label{fig_ond}
\end{figure}

The RSTN radio flux at several frequencies is plotted in Fig.~\ref{fig_rstn1}. The peaks of radio bursts  are observed within one minute before the GOES maximum, i.e. at $\sim$~09:28 UT. The flare shows 
a single peak at 0.61\,--\,2.69 GHz, whereas there are multiple peaks at 4.99\,--\,15.40 GHz frequencies. The observed radio bursts correlate well with $n$\,$\ge$\,5 detected from RESIK and with low $\delta$ derived from RHESSI. Furthermore, the bursts occurred simultaneously with a group of dm-type III bursts 
observed in the time interval 09:28:16\,--\,09:28:33 UT, as shown in the radio spectrum in Fig.~\ref{fig_ond}. Some of these bursts have positive frequency drifts and other negative ones. This indicates the electron beams moving in both directions:
downwards and upwards in the solar atmosphere. 

\subsection{2003 January 7 flare}

\begin{figure*}
\centering {\includegraphics[width=8.00cm]{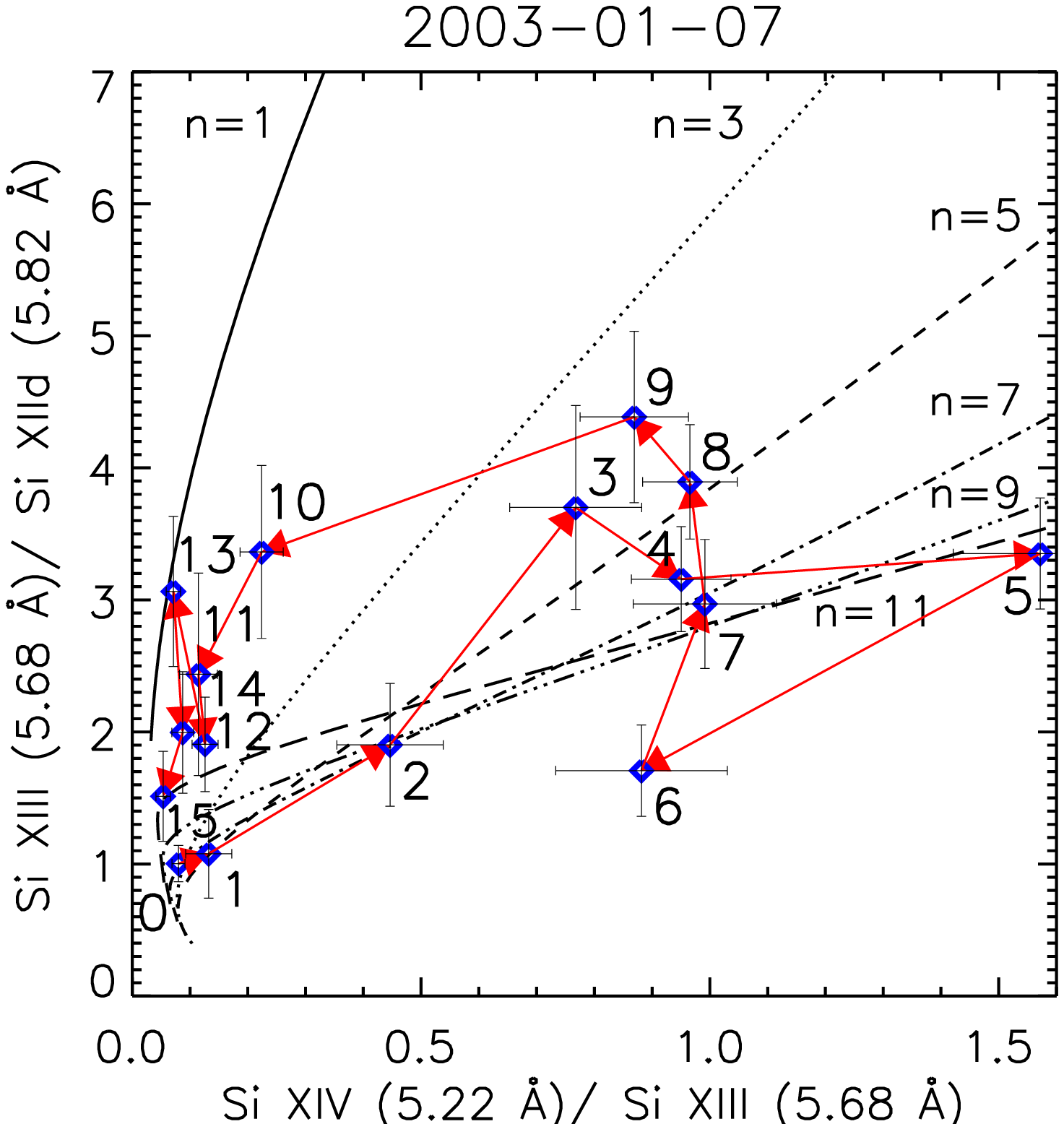} \includegraphics[width=8.00cm]{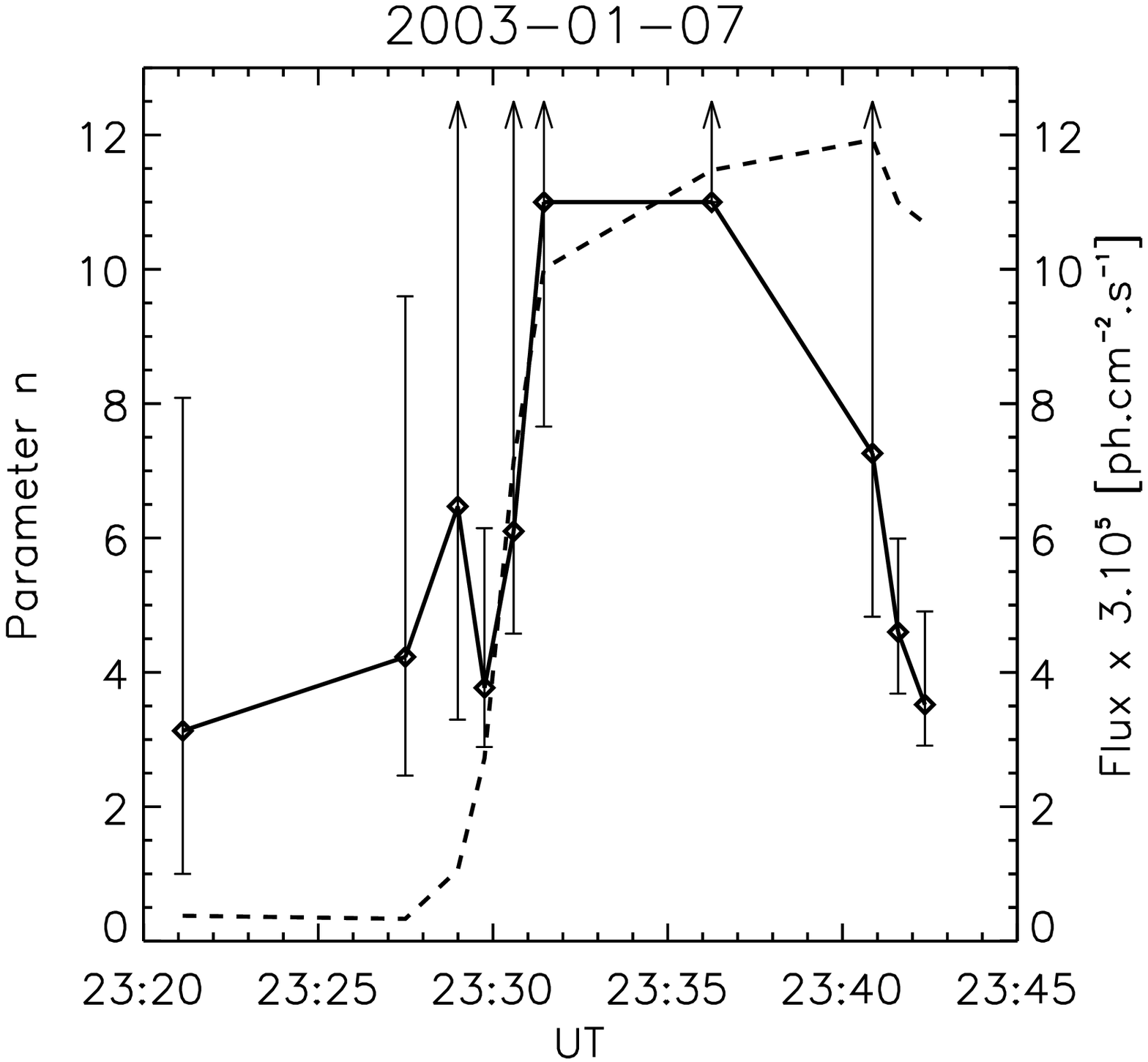}}
\centering{\includegraphics[width=8.0cm]{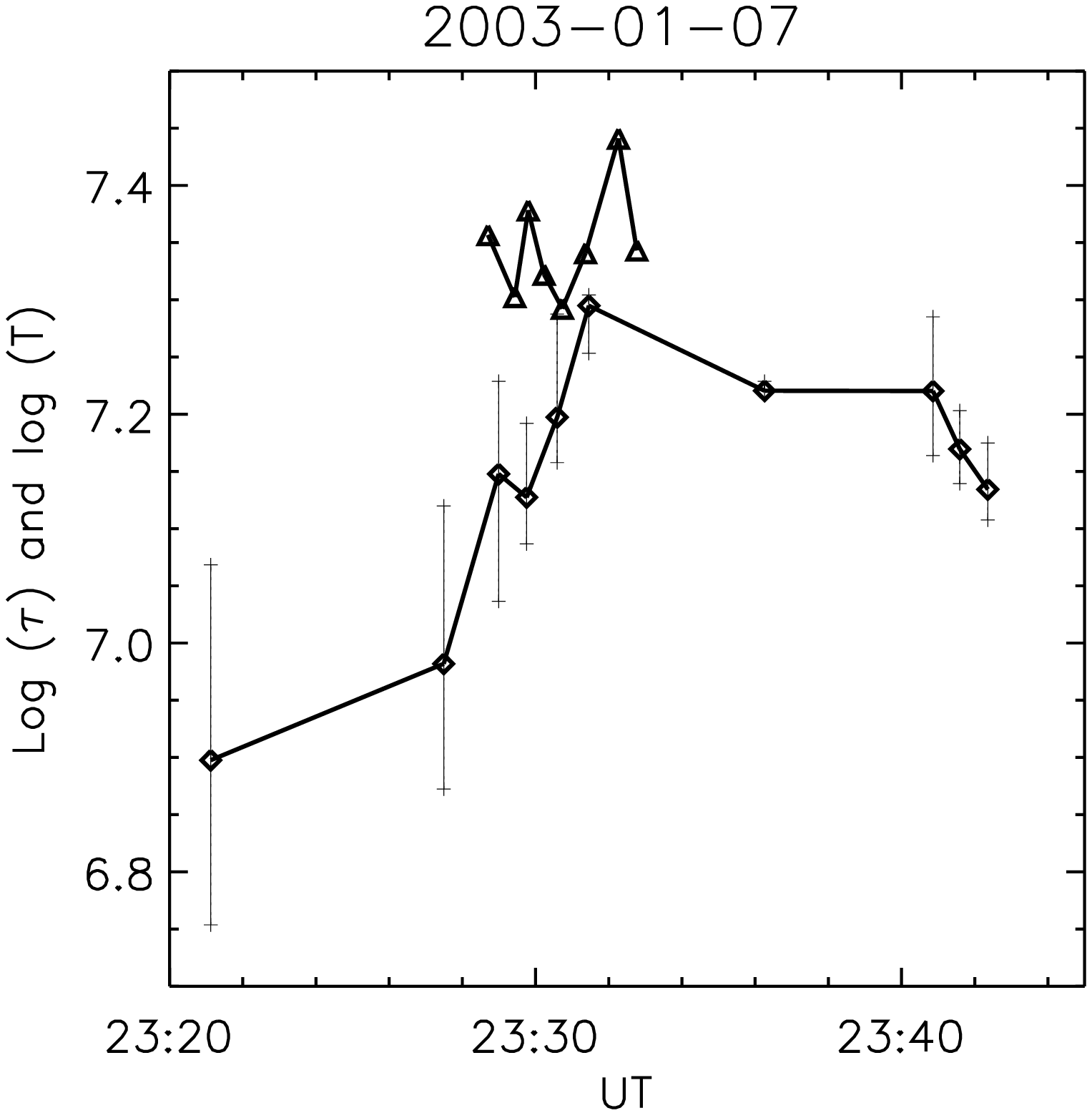}
\includegraphics[width=8.0cm]{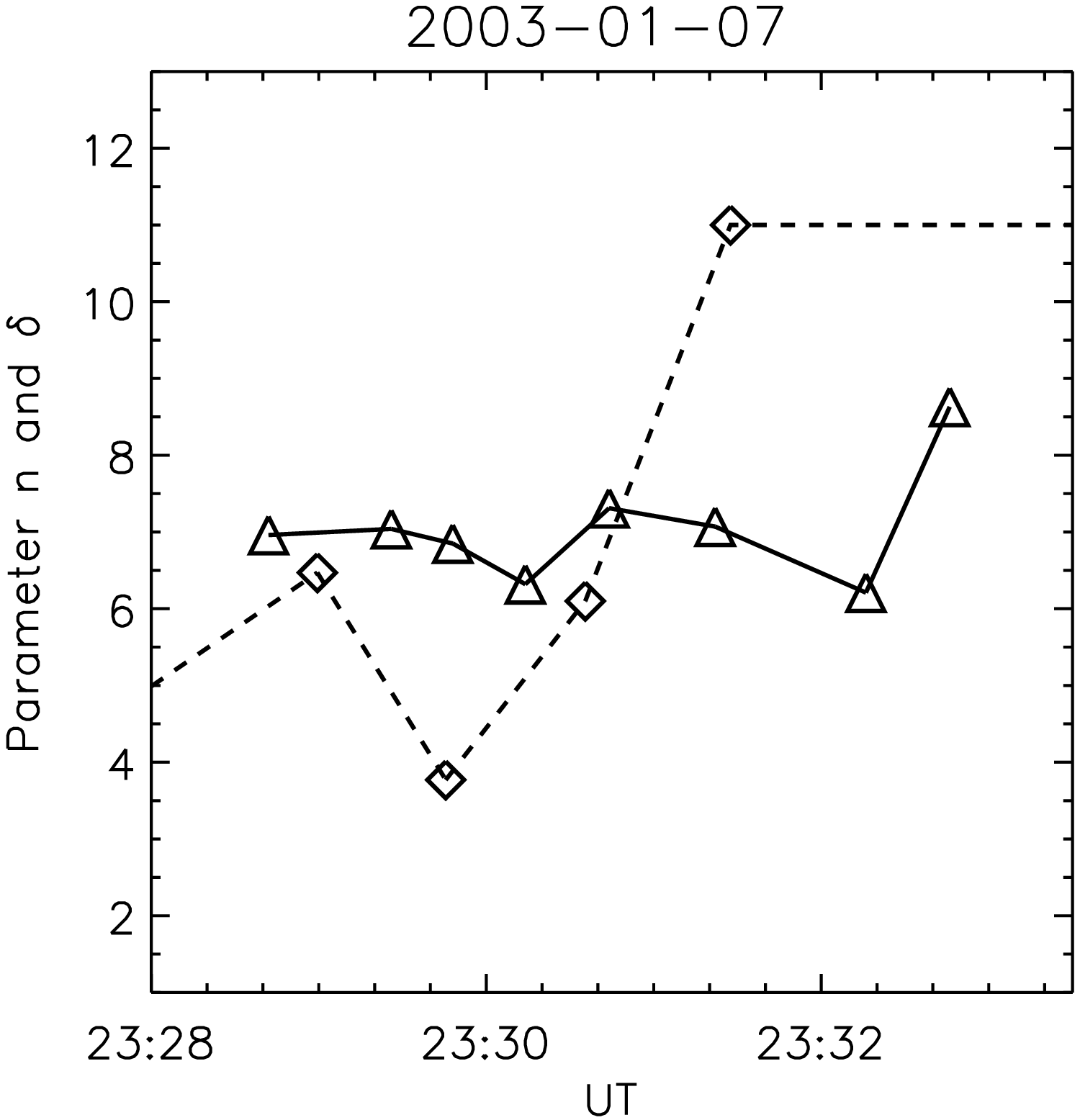}}
\caption{
2003 January 7 flare. Notation is the same as in Fig.~\ref{fig_res1}.
} 
\label{fig_res2}
\end{figure*}

This M4.9 flare occurred in the active region NOAA 10251, starting in GOES soft X-rays at 23:25 UT, reaching its maximum at 23:33 UT and ending at 23:40 UT.
During this flare, see Fig.~\ref{fig_res2}, parameter $n$  started to rise about five minutes before the GOES soft X-ray maximum and remained $\ge$\,5 until the end of the soft X-ray flare, except for a single drop at $\sim$\,23:29\,UT. After 23:42\,UT it decreased below five. The parameter $n$ generally follows the total RESIK flux 
plotted in the top right of the Fig.~\ref{fig_res2}. This is different from the previous flare where the maximum of the total RESIK flux occurred about two minutes after the maximum values of $n$. The spectral index $\delta$ shows a distinct time evolution from two other analysed flares. Its value stays almost constant, $\sim$\,7, with two noticeable drops, i.e. local minima, at 23:30\,UT and 23:32\,UT. 

\begin{figure}
\centering
\includegraphics[width=9.0cm]{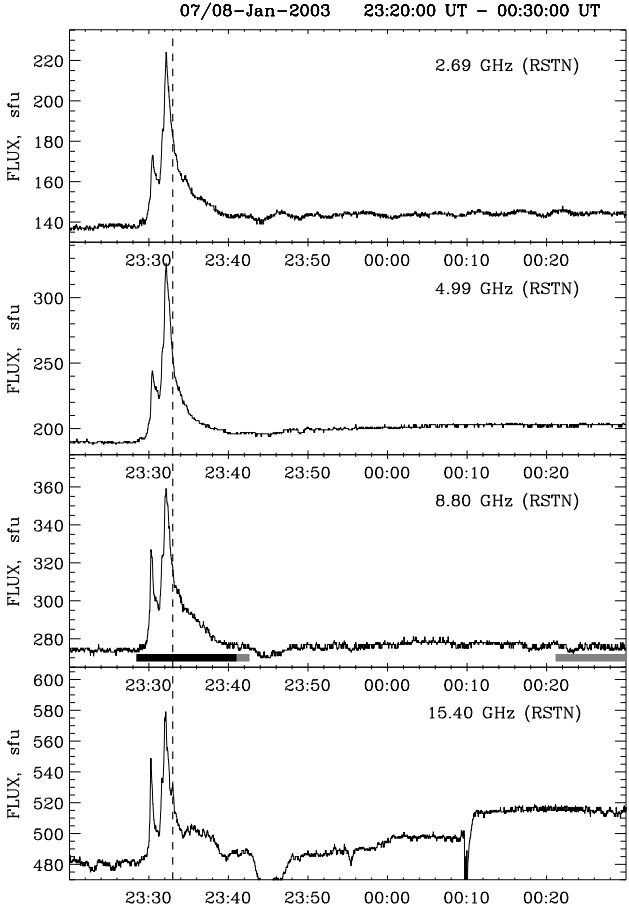}
\caption{
2003 January 7 flare. Notation is the same as in Fig.~\ref{fig_rstn1}.
}
\label{fig_rstn2}
\end{figure}

On the other hand, the radio data suggest correlations with the time behaviour of $\delta$ and $n$. The radio burst observed by RSTN (Fig.~\ref{fig_rstn2}) displays two peaks at frequencies 2.69\,--\,15.40\,GHz appearing before the soft X-ray maximum. The times of these radio peaks, i.e. $\sim$\,23:30 and $\sim$\,23:32\,UT, correspond well to the two drops in the spectral index $\delta$. Furthermore, the parameter $n$ shows an increase up to 11 at the time of the second RSTN peak and remains at that value for several minutes (see RESIK spectra 5 and 6 in Table~\ref{tab2} and Fig.~\ref{fig_res2} top right). Additionally, the enhanced radio flux is still visible up to 23:40\,UT. During that time, RESIK spectra still indicate deviations from the Maxwell distribution, exhibiting $n$\,$\ge$\,5 (Fig.~\ref{fig_rstn2}). In contrast to it, RHESSI spectra suggest thermalization of the hard X-ray emitting plasma sooner, at $\sim$\,23:33 UT, when $\delta$ rises to $\sim$\,9. Since then, no significant non-thermal component is present in the RHESSI spectra. 

Time evolutions of log\,$\tau$ and log\,$T$ during the flare are plotted in the bottom left of the Fig.~\ref{fig_res2}. The determined values of log\,$\tau$ span the range of 6.81\,--\,7.29 (0.6\,--\,1.7~keV). After reaching its maximum at $\sim$\,23:31~UT, i.e. two minutes before the soft X-ray maximum, $\tau$ starts to decrease until the end of the flare. Again, log\,$T$ obtained from RHESSI thermal component are higher than log\,$\tau$ with values between 7.29\,--\,7.44 (1.7\,--\,2.4~keV). In contrast to $\tau$, $T$ has a rather spiky character. Its maximum was observed before the soft X-ray maximum, too, at 23:32~UT. 
Tables~\ref{tab2} and \ref{tabx2} list the determined RESIK and RHESSI parameters, respectively.


\subsection{2002 October 4 flare}
The third flare was of GOES class M4.0 and occurred in the active region NOAA 10137. It started at 05:34 UT, 
reached the GOES maximum at 05:38 UT, and ended at 05:41 UT.
For this flare the parameter $n$\,$\ge$\,5 was observed between $\sim$~05:35\,--\,05:42~UT (Fig.~\ref{fig_res3}, top right). 
After $\sim$ 05:39~UT $n$ started to decrease. Like the 2003 January 7 flare, the global behaviour of 
$n$ follows
the RESIK total flux. On the other hand, the time evolution of the spectral index $\delta$ (Fig.~\ref{fig_res3}, bottom right) has a similar shape 
to the 2003 February 22 flare (Fig.~\ref{fig_res1}, bottom right), but in this case the RHESSI spectra are much steeper, therefore $\delta$ has higher values. 
Fast variations of $n$ and its large uncertainties  during the rise and the maximum phase are not a real feature, 
but are probably due to low photon flux and related high statistical errors.
The time evolution of log\,$T$ obtained from RHESSI is nearly constant, 
and its values are again higher than $\log\tau$. 
Tables~\ref{tab3} and \ref{tabx3} list the determined RESIK and RHESSI parameters, respectively.

Radio burst observed by RSTN (Fig.~\ref{fig_rstn3}) occurred before the flare's soft X-ray maximum and started at $\sim$~05:36 UT. The shape of the radio flux shows a double peak at frequencies 4.99~GHz and 8.80~GHz with the first peak occurring at about 05:36~UT and the second one about 05:37~UT. At the highest RSTN frequency, 15.40 GHz corresponding to the lower chromosphere, the burst shows a triple peak between 05:36\,--\,05:37~UT. In contrast to this, the 2.69~GHz flux shows 
a single broad peak with a maximum at about 05:37~UT. There radio bursts correlate well with the presence of the RHESSI non-thermal component observed at 05:35:28\,--\,05:37:32~UT. Furthermore, enhanced radio flux at all four frequencies, as well as n-distribution of $n\ge5$, were still visible after the GOES  maximum.

\section{RHESSI and n-distribution}
\label{sect:rhessi_ndistr} 

If there is any non-thermal n-distribution really present in flaring plasma, its electrons should also contribute to the production of bremsstrahlung radiation. Therefore, we included the thin-target model of bremsstrahlung from n-distribution into fitting of RHESSI spectra and tried to test the parameters of n-distribution independently from RESIK. We performed such an analysis only for the first two flares, 2003 February 22 and 2003 January 7, as we needed to use the RHESSI spectra observed in A0 attenuator state for this purpose. This allowed us to examine the lowest part of the RHESSI energy range that is closest to the energies observed by RESIK. Unlike analysis of the RESIK X-ray line spectra, we suppose that RHESSI will give us the information about continuum radiation produced by the n-distribution.

We therefore revisited A0 RHESSI fits, as seen in the first row in Tables ~\ref{tabx1} and \ref{tabx2} and Fig.~\ref{fig_rhessi_o}, where the fits 
are shown with their normalized residuals. Both fits started at 6 keV and they had $\chi_{\nu}^{2}$\,$\sim$\,1 ($\chi_{\nu}^{2}$ means the reduced $\chi^{2}$). Then, using the best-fit parameters as a starting point, we pushed the low-energy boundary down to 4 keV and let the parameters of the isothermal and the power-law component vary.(When either {\it pileup\_mod} or {\it drm\_mod} pseudo-functions were used, the values of their parameters were taken from the previous fits and kept fixed.) However, no satisfactory fits could be obtained in this way. In this way $\chi_{\nu}^{2}$>5 was found for both flares,
and the residuals were unacceptably large, showing a systematic pattern below 10~keV (Fig.~\ref{fig_rhessi_w}).

Therefore, in the last set of fits we added the thin-target bremsstrahlung radiation from the n-distribution. This component was calculated by a new fitting function, {\it thin\_ndistr}, which has the four parameters to fit: $n$,~$\tau$, "emission measure" (i.e. ambient proton density $\times$ volume $\times$ electron density in the n-distribution), and a high-energy cutoff. (The function, {\it thin\_ndistr}, was incorporated into SSW tree and can be accessed from OSPEX.) The high-energy cutoff for the n-distribution is introduced for a numerical purpose only. It was set to 1 MeV and fixed for all measurements. The initial values of other three parameters of the {\it thin\_ndistr} function were set to those obtained from the RESIK diagnostics. 

These fits (Fig.~\ref{fig_rhessi_n}) improved significantly both in terms of residuals and $\chi_{\nu}^{2}$. The component due to the n-distribution describes the lowest energies of the fits, whereas the spectrum at higher energies is not affected, and its parameters are similar to those from the original fits starting from 6 keV. Tables~\ref{tabx4} and~\ref{tabx5} summarize the parameters for individual flares and for all three fits: 
the original fit, the fit with the enlarged fitting range, and the one combined with the n-distribution bremsstrahlung. To complete the derived parameters, we also list RESIK results for the corresponding time intervals in those tables. Tables~\ref{tabx4} and~\ref{tabx5} and Fig.~\ref{fig_rhessi_n} show the fits for a specific case of the parameter $n$ being fixed at a value derived from RESIK. The reason is that the studied Si lines provide the best possibility for determining the parameter $n$ because the line ratios with the satellite lines are very sensitive to the shape of the distribution considered. In addition to the fits with fixed $n$, we also analysed fits setting $n$ as a free parameter. Both sets of fits were similar in terms of $\chi^2$ and residuals. However, in the latter case the best fit was not well defined since a wide range of n-distribution parameters resulted in a similar value for $\chi^2_\nu$. Such behaviour suggested that the parameter uncertainties could be large.

In order to estimate the uncertainties of the fitted parameters, we examined the $\chi^2$ values by keeping a selected parameter fixed within a range of values and by fitting the other parameters (see also {\it chi2\_map} method in OSPEX). Having done so, we found out that $\chi^2$ does not exhibit well localized minima for n-distribution parameters. Moreover, in some cases only their lower or upper limits could be estimated. This is a situation similar to the low-energy cutoff (and consequently the total flux) for which it is known that in some cases the spectral fits could only provide an upper limit estimate. A cutoff below the upper limit still fits the data because at low energies the non-thermal flux is usually well below the thermal component \citep{sui05}. On the other hand, this  $\chi^2$ mapping technique allowed us to obtain thermal parameters and spectral index $\delta$ uncertainties, i.e. as values corresponding to $\chi^2_\mathrm{min}+1$. However, we note that this approach does not provide joint uncertainties of fitted parameters, but instead 
corresponds to a single parameter uncertainty independent of any knowledge about the others. To obtain a joint estimate of several parameters at a 68$\%$ ("1$\sigma$") confidence level, a much larger region, $\chi^2_\mathrm{min}+ \Delta\chi^2$, would have to be investigated \citep{lam76,naka10}. 
But this is beyond the scope of the paper.

The estimations of the parameter uncertainties are given in Tables~\ref{tabx4}  and~\ref{tabx5}. For the case of the 2003 February 22 flare, the fit with $n$ fixed at the RESIK value, i.e. $n$\,=\,11, gives n-distribution parameters that are slightly above the estimated $1\sigma$ range, i.e. $n$=2\,--\,9, $k\tau$\,=\,(1.0\,--\,1.5)\,keV (but well within $2\sigma$). However, due to the weak dependence of $\chi^2$ on the n-distribution parameters, such a fit could be considered acceptable. Although the uncertainties of $n$ and $\tau$ obtained either from RHESSI or RESIK data are large for both events, their ranges overlap. 

We also performed a test to check the reality of the excess emission detected in the lowest RHESSI energies.
A spectrum from a different detector (3F) was analysed and resulting fitting parameters agreed well within the uncertainties with those
discussed above. Therefore, the excess emission, fitted here by the n-distribution, is not an instrumental artefact but a real feature.

Using the reconstructed images, we estimated the electron densities in plasma for different electron distributions. From the RHESSI fits we have the information about the emission measure for the thermal component ($EM_\mathrm{th}$) and for the n-distribution ($EM_\mathrm{n}$), as well as the total rate of injected electrons ($F_\mathrm{T}$) for the power law. The n-distribution appears at the lowest part of the RHESSI spectrum. We therefore  reconstructed the images in the 3\,--\,6~keV range for it, the 6\,--\,12~keV channel was used for the thermal component, and the 25\,--\,50~keV channel was dedicated to the hard X-ray emission. From the images we determined the area $S$ within the 70\% contour of the maximum and supposed that the particular volume of emitting plasma has a spherical shape with a circular cross section $S$. Both for the n-distribution and 
the Maxwell distribution we supposed that $N_\mathrm{p}=N_\mathrm{e}$ and $EM_\mathrm{th}=(N^\mathrm{th}_\mathrm{e})^{2}V_\mathrm{th}$ and $EM_\mathrm{n}=({N^\mathrm{n}_\mathrm{e}})^{2}V_\mathrm{n}$, where index `$\mathrm{th}$' is for the thermal component and `$\mathrm{n}$' corresponds to the n-distribution. The 70\% contours in the 3\,--\,6~keV and 6\,--\,12~keV channels were almost the same (Fig.~\ref{fig_rhessi_imgs}), so we used only the 
6\,--\,12~keV channel for the electron density estimation, 
assuming $V_\mathrm{th}=V_\mathrm{n}$. The electron density for the power-law component was derived using the relation $N^\mathrm{PL}_\mathrm{e}$\,$\sim$\,$F_\mathrm{T}/(S_\mathrm{hard}\sqrt{2E_\mathrm{c}/m_\mathrm{e}})$, where $S_\mathrm{hard}$ is the area in the 25\,--\,50~keV hard X-ray channel. 

In both flares the electron densities for n-distribution are three to four times higher than the electron densities of the thermal component. Moreover, they are about three to four orders of magnitude higher than the electron densities that correspond to the power-law distribution.
Furthermore, the thermal and n-distribution electron densities are similar in both events, while $N_\mathrm{e}$ for the power-law distribution differs nearly by one order. The 2003 January 7 flare was a limb flare (Fig.~\ref{fig_rhessi_imgs}, right), so we register only its projection. It is likely that the flare footpoints, from where the dominant contribution to the 25\,--\,50 keV channel comes, are the most affected by that projection.

\section{Discussion}

Basically, we suppose that a non-thermality in soft X-ray range manifested by the n-distribution is connected with a non-thermality in the hard X-ray region. Then at a first sight, one could expect to see an `X-pattern' in  $n$ and $\delta$ plots, i.e. when $\delta$ is the lowest, $n$ reaches the highest values. 
But, in
reality this connection is not so simple and does not depend on relation between $n$ and $\delta$ only. It very likely also depends on electron flux density \citep{dzifkar08} if we assume that the n-distribution is produced because of the return current. 
None of  the three analysed flares exhibits a clear `X-pattern'. There is a hint of such tendency, i.e. a smooth decrease in $n$ versus a rise 
in $\delta$, 
for the simple compact flare from February 22, 2003 of the GOES class C (see Fig.~\ref{fig_res1}, bottom right), however 
the large uncertainties of $n$ do not allow us to consider this as significant.
The other two M class flares were more complex. The 2002 October 4 flare was a two ribbon flare occurring near the limb centre. The 2003 January 7 flare was a limb flare 
exhibiting a rather complex magnetic configuration, possibly with a plasmoid structure \citep[Fig.~\ref{fig_rhessi_imgs} and][Fig.~2]{dzi08}.
Additionally, both these flares show longer time occurrence of the non-thermal n-distribution than 
the power-law component observed in RHESSI spectra.

In all three flares the time interval with $n$\,$\ge$\,5 corresponds well with radio bursts observed at frequencies 4.99\,--\,8.80 GHz. Also, we found a good correlation of radio emission with hard X-rays related to electron beams. This behaviour is known and has been studied, e.g., by \cite{aschw95}, \cite{Benz05}, and \cite{fk07}. For the 2003 February 22 flare, no enhanced radio emission was observed after the burst. On the other hand, the long duration enhanced radio emission was presented, together with $n\ge5$ for both M class flares, but a correlation with $\delta$ shows a shorter duration.

\begin{figure*}
\centering {\includegraphics[width=8.0cm]{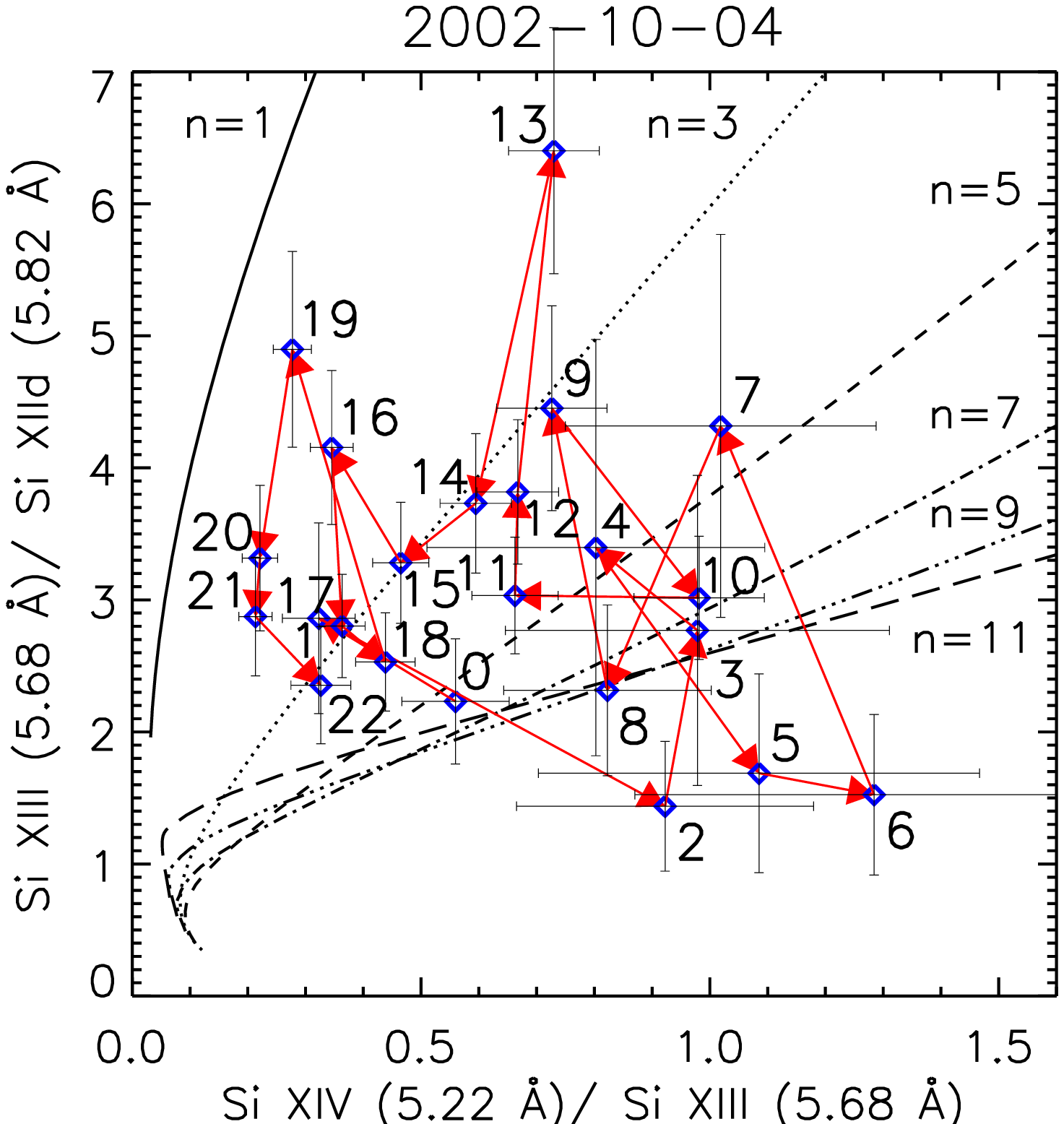}\includegraphics[width=8.0cm]{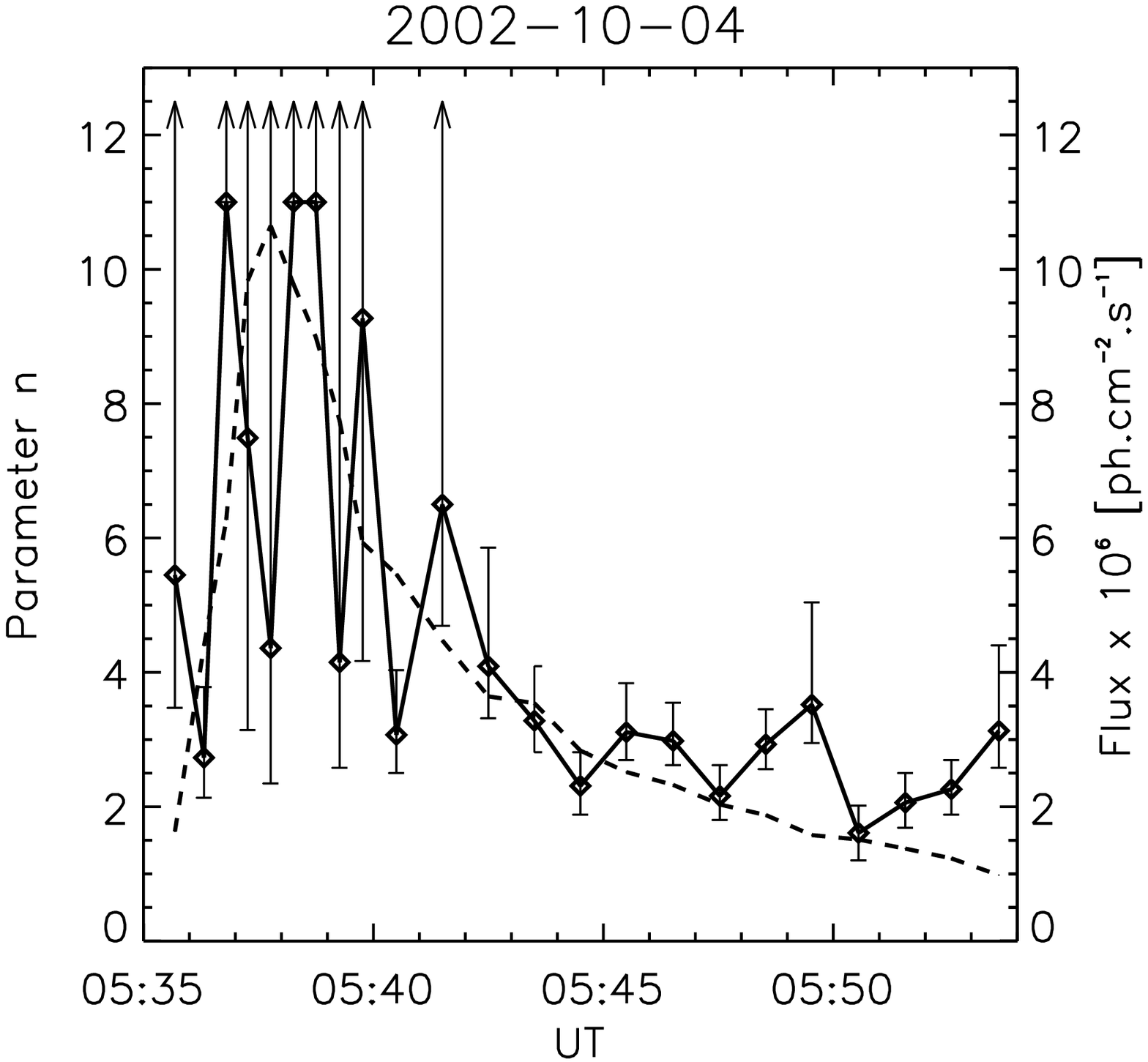}}
\centering{\includegraphics[width=8.0cm]{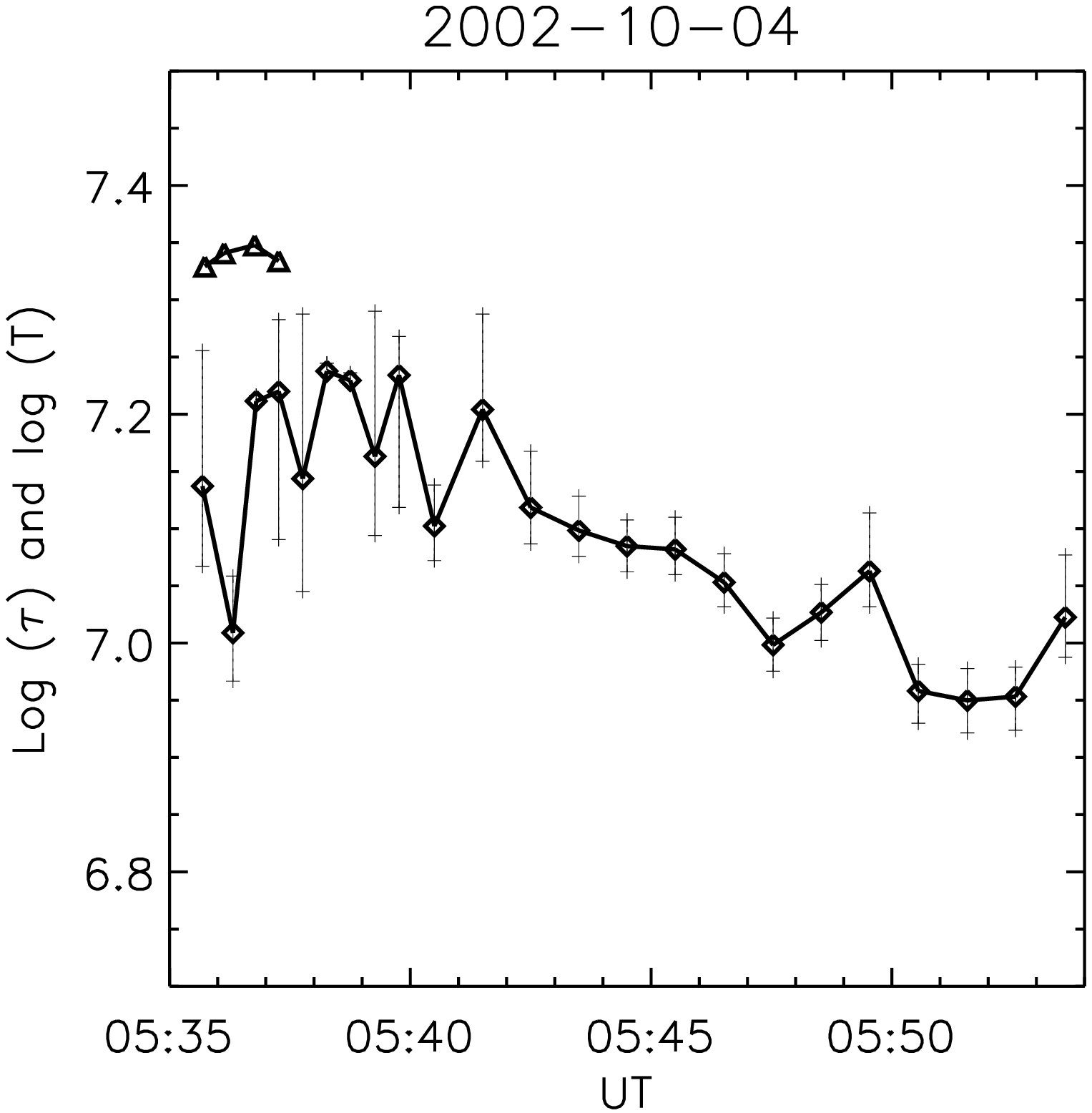}\includegraphics[width=8.0cm]{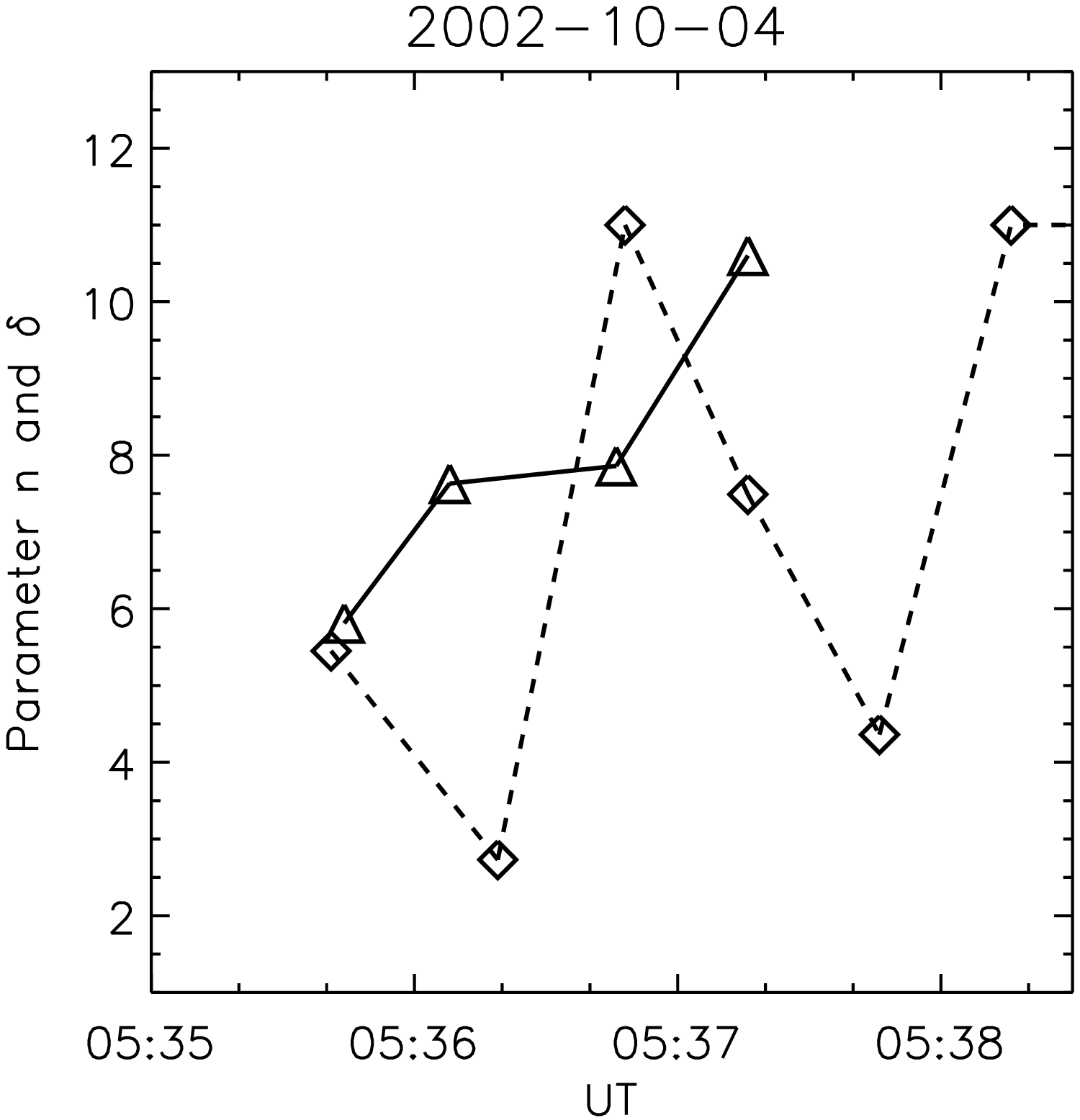}}
\caption{
2002 October 4 flare. Notation is the same as in Fig.~\ref{fig_res1}.
} 
\label{fig_res3}
\end{figure*}

\begin{table}
\centering
   \caption{Derived electron densities for Maxwell ($N^\mathrm{th}_\mathrm{e}$), power law ($N^\mathrm{PL}_\mathrm{e}$), and n-distributions ($N^\mathrm{n}_\mathrm{e}$)}. 
    \renewcommand{\arraystretch}{1.2}
\begin{center}
    \begin{tabular}[h]{ccc}
      \hline \hline
      Distribution & February 22, 2003 & January 7, 2003\\ \hline
      $N^\mathrm{th}_\mathrm{e}$ [cm$^{-3}$] &  1.6$\times$10$^{10}$ & 9.8$\times$10$^{9}$ \\ 
      $N^\mathrm{n}_\mathrm{e}$ [cm$^{-3}$] & 4.0$\times$10$^{10}$ & 3.5$\times$10$^{10}$ \\
      $N^\mathrm{PL}_\mathrm{e}$ [cm$^{-3}$] & 4.7$\times$10$^{6}$ & 2.8$\times$10$^{7}$  \\
      \hline \\
      \end{tabular}
      \end{center}
    \label{tab_ne}
\end{table}

For the 2003 February 22 flare, the type III bursts with positive and negative frequency drifts were observed within the time interval with $n$\,$\ge$\,5. These bursts indicate the electron beams are moving downwards and upwards in the solar atmosphere. Because each electron beam needs to be accompanied also by the return current electrons \citep{kar09}, such a radio spectrum can be considered as evidence of the non-Maxwell electron distribution function representing an electron beam with a return current in the region of the radio source. Assuming radio emission generated on the plasma frequency 
(800\,--\,4500~GHz, see Fig.~\ref{fig_ond}), 
these type III bursts thus show that deviations from the Maxwell distribution are present in the atmospheric layers with the plasma density 
$N_\mathrm{e}$
from 7.9$\times$10$^{9}$ cm$^{-3}$ to 2.5$\times$10$^{11}$ cm$^{-3}$, i.e. in the chromosphere through the transition layer to the low corona. The electron density of the n-distribution determined from the RHESSI spectral fits and images, see~Table~\ref{tab_ne}, agrees well with the above estimate from the radio emission. This further supports the idea of electron beams and of an accompanying return current that causes the n-distribution. 
Additionally, assuming that the ratio  $N^\mathrm{III}_\mathrm{e}/N_\mathrm{e}= 10^{-6} - 10^{-7}$ \citep[][p.~203~-~204]{mel80}, where $N^\mathrm{III}_\mathrm{e}$ is the plasma density of the electron beam generating type III bursts,
then we can see that $N^\mathrm{III}_\mathrm{e}$ is much lower, $\approx 7.9\times 10^{3}-2.5\times10^{5}\,\mbox{cm}^{-3}$, than
the plasma densities of the n-distributions presented in Table~\ref{tab_ne}. Thus a contribution of the
electron beam electrons producing type III bursts to the n-distribution is very small.

\begin{figure}
\centering \includegraphics[width=9.0cm]{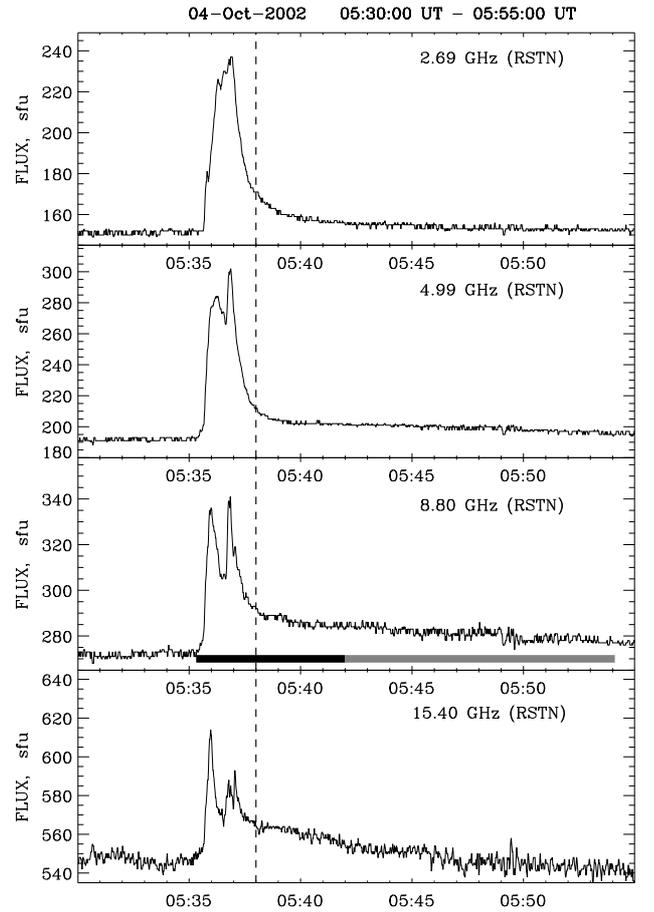}
\caption{
2002 October 4 flare. Notation is the same as in Fig.~\ref{fig_rstn1}.
}
\label{fig_rstn3}
\end{figure}

Furthermore, we made the following comparison of parameters of flares in February 22, 2003 and January 7, 2003. In the model in which the n-distribution is formed by electrons of the return current induced by the
electron beam \citep{dzifkar08}, we can write 
$N_\mathrm{e}^\mathrm{n}v_\mathrm{RC}^\mathrm{n} = N_\mathrm{b} v_\mathrm{b},$ where $N_\mathrm{e}^\mathrm{n}$ is the 
electron density corresponding to the n-distribution,
$N_\mathrm{b}$ the density of the power-law tail (electron beam), $v_\mathrm{RC}^\mathrm{n}$ the mean return current velocity in the proton coordinate system, and $v_b$ the mean velocity of the power-law tail (beam). The 
n-distribution densities in both the flares are nearly the same (Table ~\ref{tab_ne}). Therefore, if we assume the same velocity $v_\mathrm{RC}^\mathrm{n}$, i.e. the same n-distributions in both flares, then the mean velocity of the power-law tail (electron beam) $v_b$ in the 2003 February 22 flare should be greater than that in the 2003 January 7 flare. However, estimation of the mean velocities significantly depends on the low-energy cutoff that was determined with large uncertainty. Nevertheless, the power-law index $\delta$\,$\sim$\,3 in the 2003 February 22 flare is much smaller than that ($\delta$\,$\sim$\,6.5) in the 2003 January 7 flare, which means that the mean velocity in the 2003 February 22 flare is greater than in 2003 January 7, as estimated. That also supports the model of the n-distribution based on the return current.

We expected that  the bremsstrahlung from n-distribution would very likely manifest itself at low energies of the RHESSI spectral window. This was confirmed by the RHESSI fits of n-distribution discussed in the previous section. We note that in both cases the n-distribution significantly contributes to the X-ray spectrum in a rather narrow energy range, therefore the determined parameters of the n-distribution have large uncertainties. In all spectra the thermal component had to be considered to account for the emission in the $\sim$ 6\,--\,10~keV energy range. Forcing the n-distribution component to higher energies instead of using the thermal component did not work: fits consisting of an n-distribution, two Gaussian lines (as approximations of the line complexes at $\sim$\,6.7 and 8 keV), and a power-law high-energy tail did not lead to acceptable results.
Finally, we also fitted the RHESSI spectra using a double power law and an isothermal component, i.e.
without the n-distribution, to examine a possibility that the low-energy emission is due to a flat power-law
component extending down to keV energies. Although an acceptable fit of $\chi^2_\nu\sim 1.3$ was 
obtained, the resulting low-energy power-law part was very steep; i. e., $\delta_{\rm L}\sim 11$ with 
$\delta_{\rm L} > \delta$, thus resembling the n-distribution component.

Neither RHESSI nor RESIK spectra are spatially resolved. The radiation produced by the thermal component and the n-distribution comes from the coronal plasma, while hard X-ray bremsstrahlung origins primarily at the footpoints (Fig.~\ref{fig_rhessi_imgs}). The 25\,--\,50~keV contour is the largest one with 3\,--\,6~keV and 6\,--\,12~keV countours in the middle (Fig.~\ref{fig_rhessi_imgs} , right) or in its upper part (Fig.~\ref{fig_rhessi_imgs}, left). Even though the contours of the 3\,--\,6 keV and 6\,--\,12 keV channels overlay each other, we suppose that the n-distribution and the thermal one are localized in different substructures of flaring plasma. There are several reasons that led to this conclusion.

\begin{figure*}
\centering {\includegraphics[angle=90,width=9.2cm]{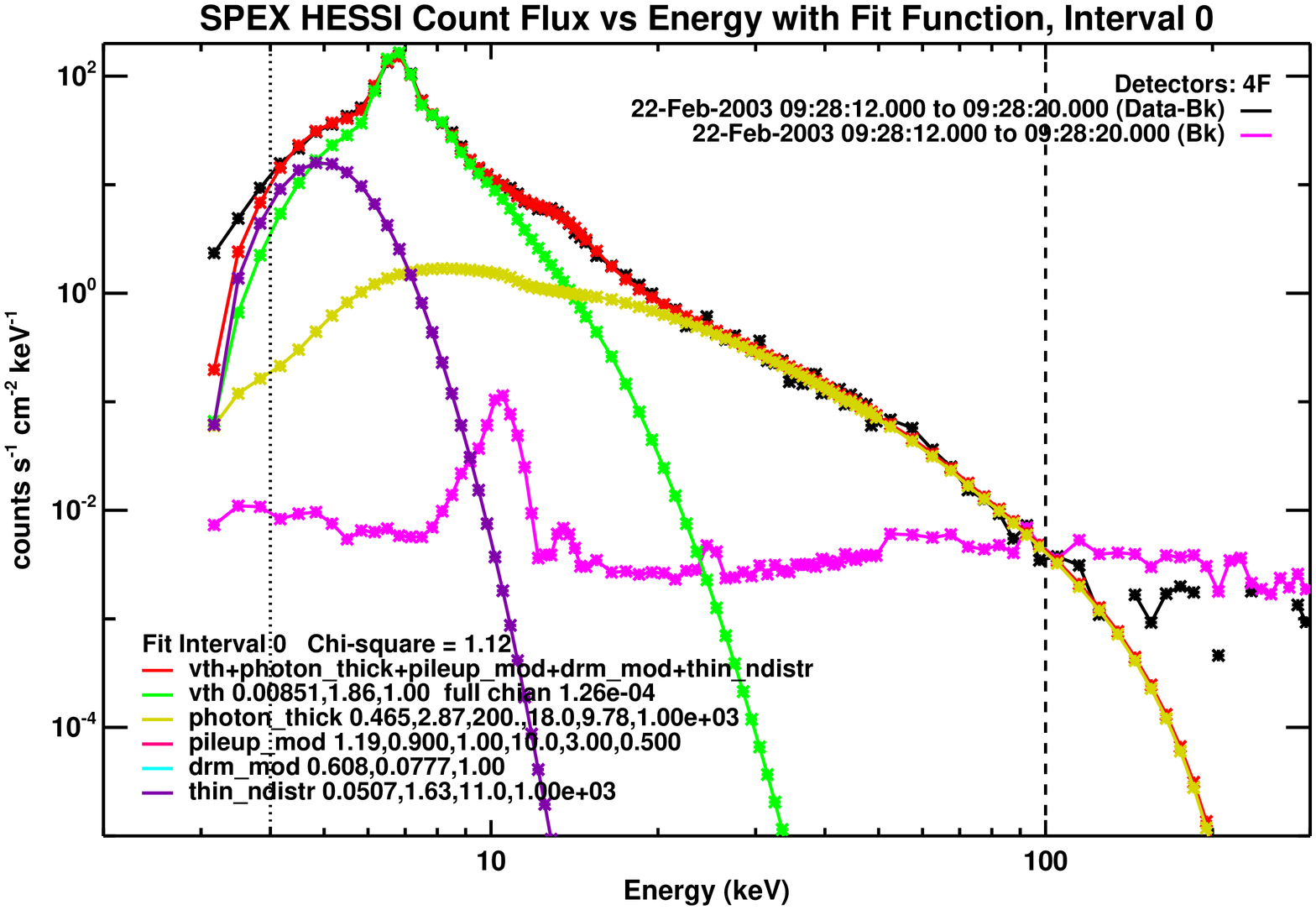}\includegraphics[angle=90,width=9.2cm]{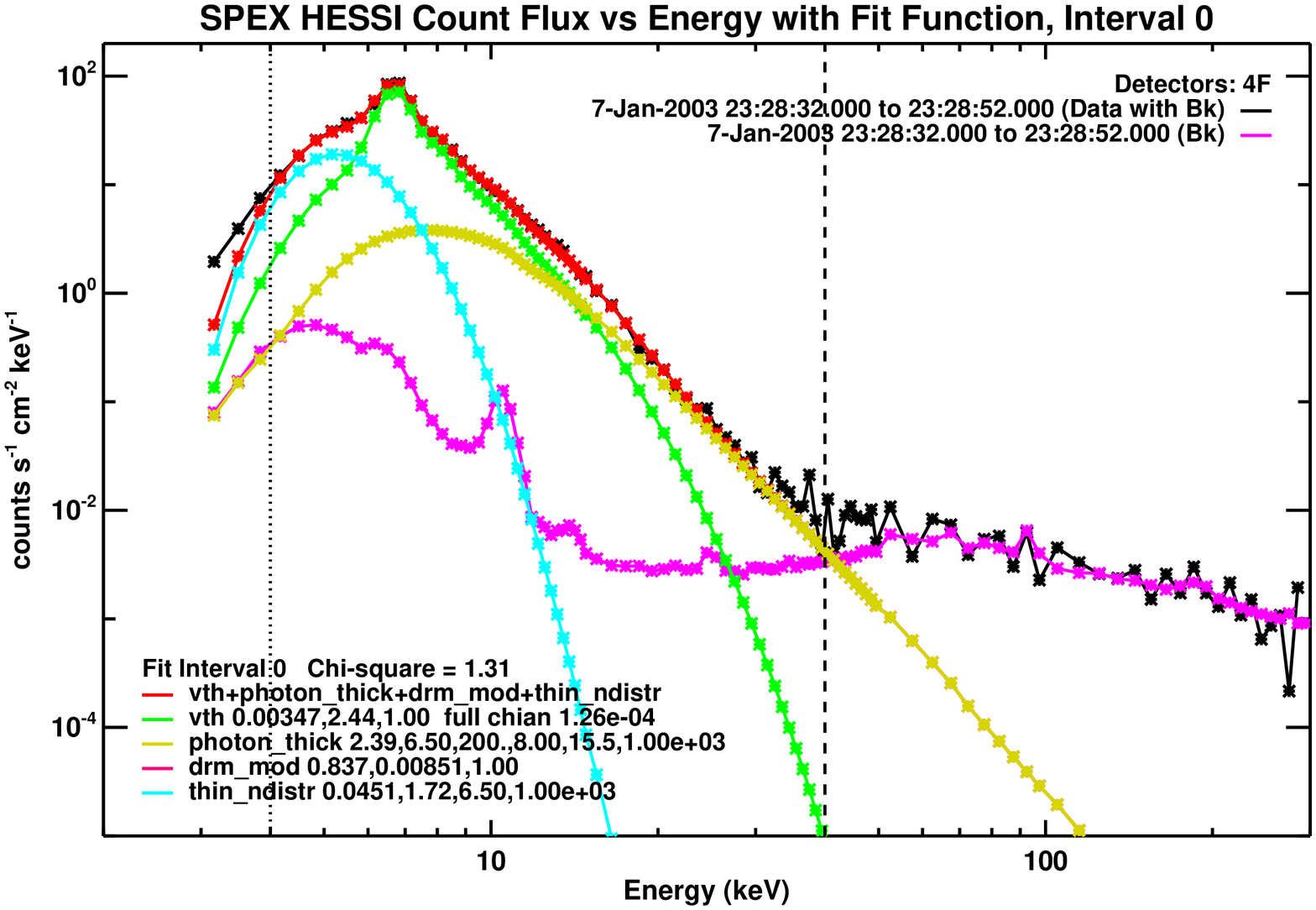}}
{\includegraphics[angle=90,width=9.2cm]{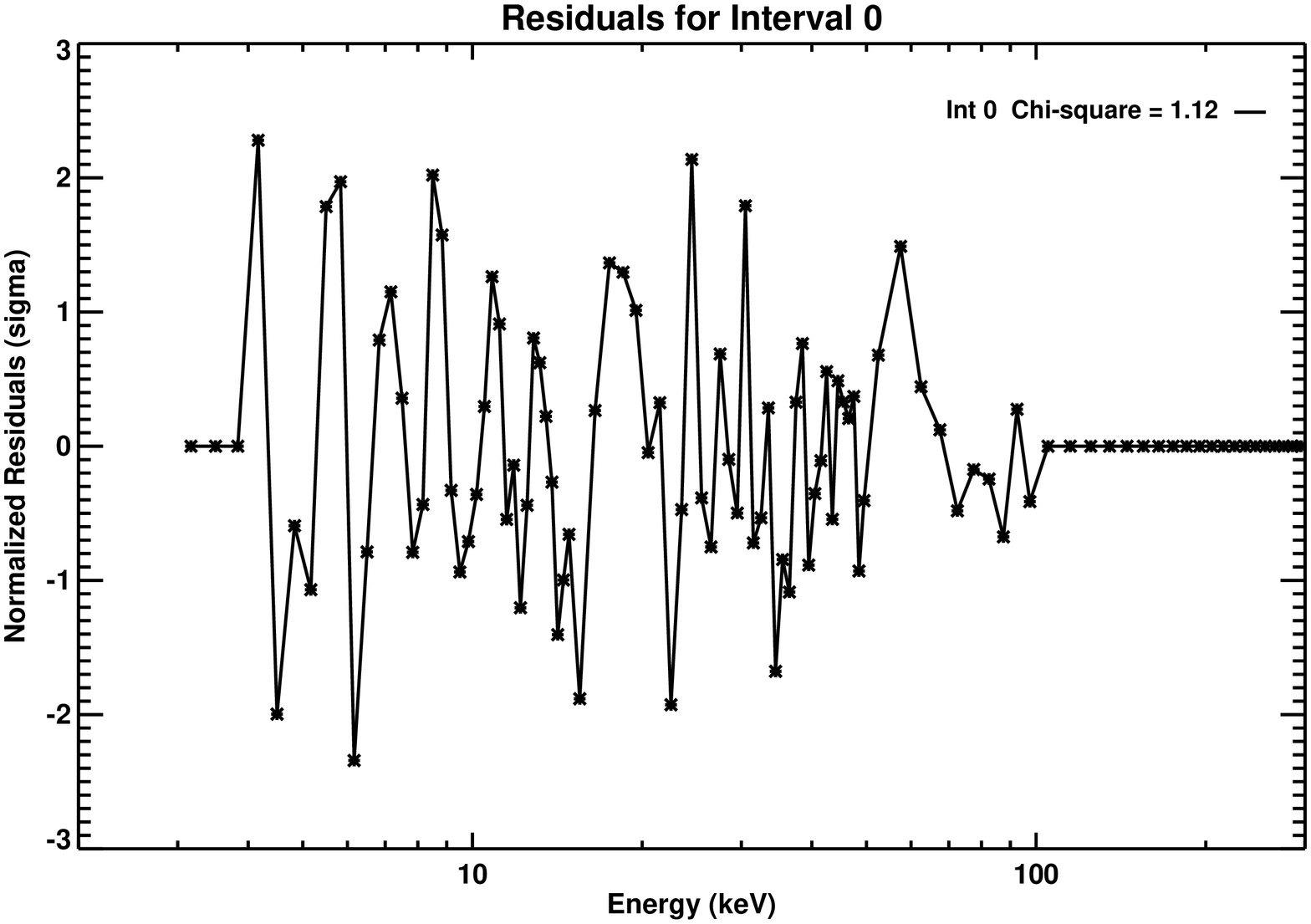}\includegraphics[angle=90,width=9.2cm]{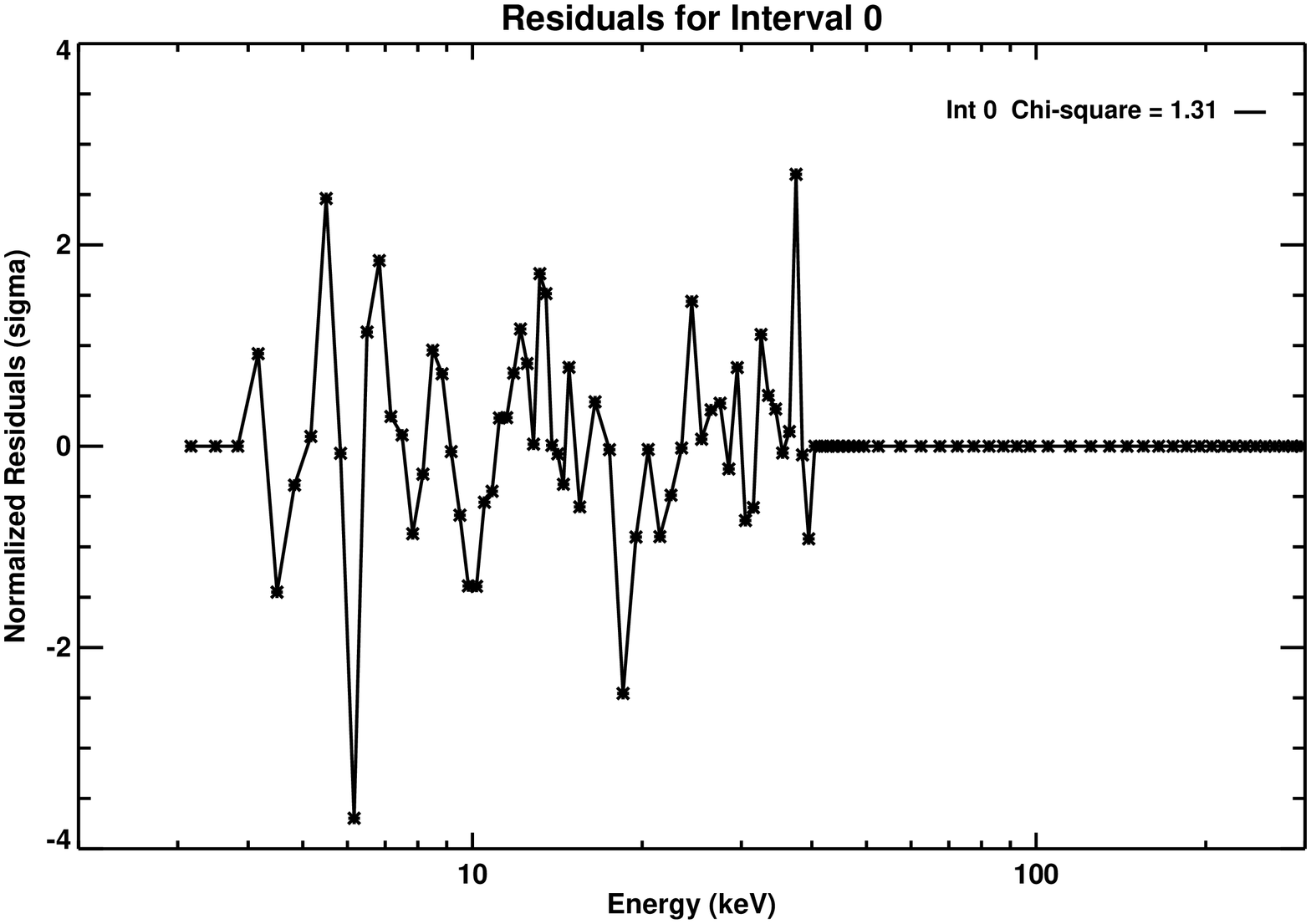}}
\caption{RHESSI fits (from 4 keV) of the single power law (thick-target model), the thermal component, and  the n-distribution (thin-target model) in 
the low energy range, together with residuals: the 2003 February 22 flare (left) and the 2003 January 7 flare (right).}
\label{fig_rhessi_n}
\end{figure*}

\begin{figure*}
\centering {\includegraphics[width=9.2cm]{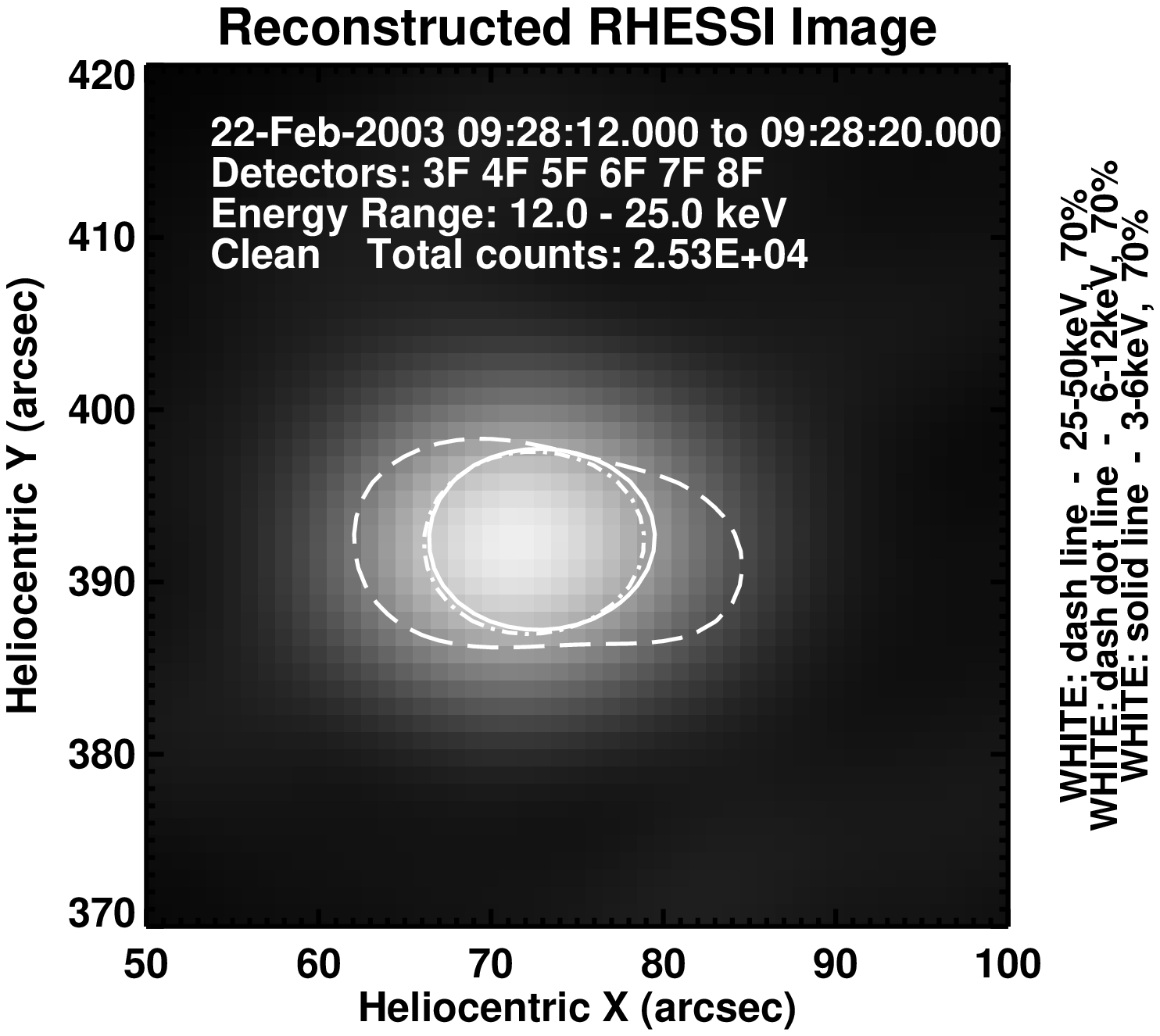}\includegraphics[width=9.2cm]{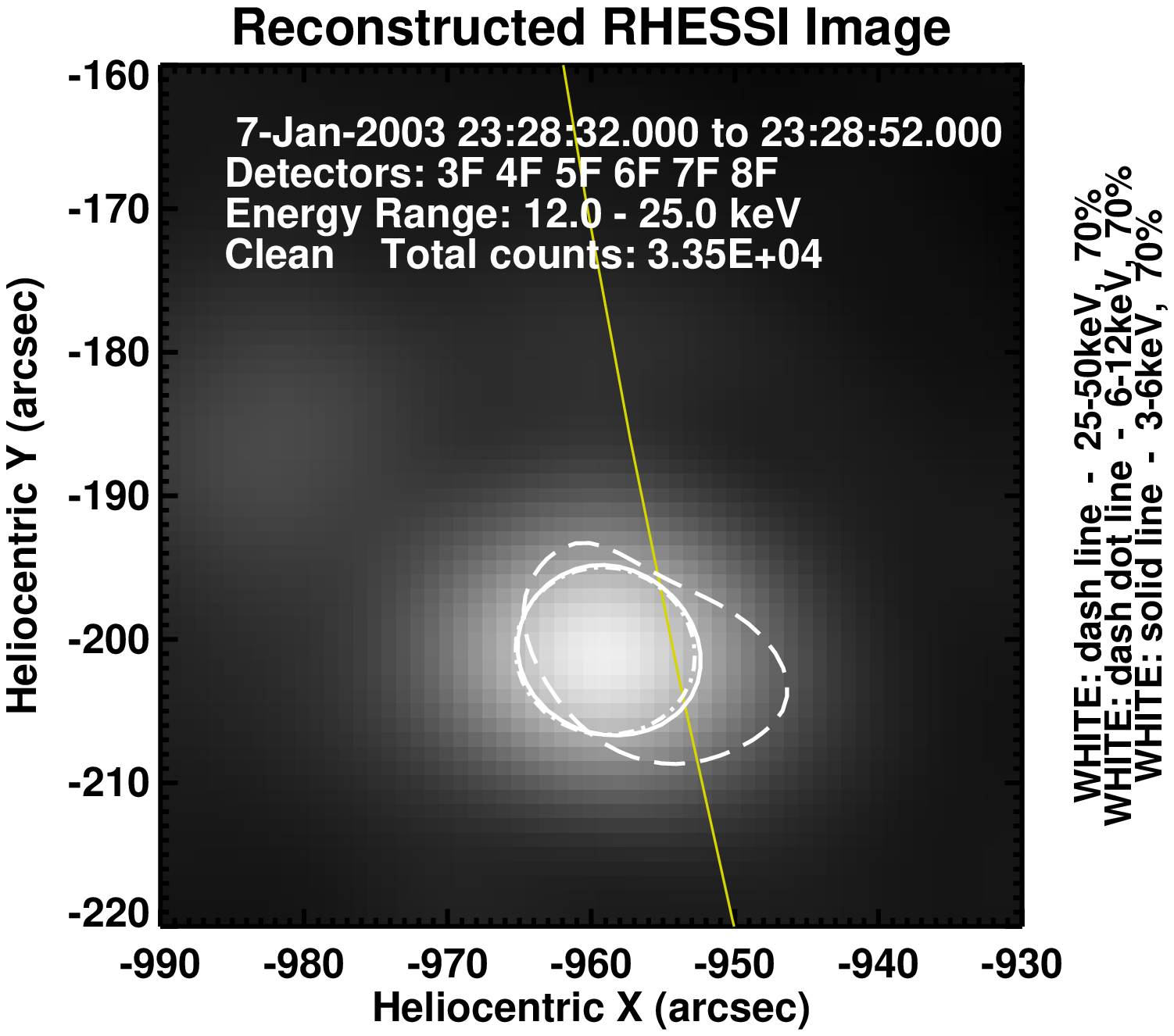}}
\caption{RHESSI images in the 12\,--\,25 keV channel with contours of 70\% from channels: 3\,--\,6 keV (full line), 6\,--\,12 keV (dash-dot line), and 25\,--\,50 keV (long dash line): the 2003 February 22 flare (left) and the 2003 January 7 flare (right). The thin solid line marks the solar limb for the 
2003 January 7 flare.}
\label{fig_rhessi_imgs}
\end{figure*}

First, $T$ determined from the RHESSI thermal component is always higher than the values of $\tau$ from the n-distribution. This difference between log\,$T$ and log\,$\tau$ is usually 0.1\,--\,0.2 dex, and it cannot be explained by the measured uncertainties (Figs.~\ref{fig_res1},~\ref{fig_res2},~\ref{fig_res3}, bottom left). The enhanced intensities of \ion{Si}{xii}d lines observed by RESIK are the second argument for a different spatial location of the Maxwell and the n-distribution. The high ratio of \ion{Si}{xii}d 
to \ion{Si}{xiii} lines needs a strong gradient of particles in the energy range above approximately 1.5 keV (Figs.~\ref{fig_ndistr},~\ref{fig_sch}). The high number of electrons with the energy corresponding to the excitation energy of doubly excited state results in the high intensities of the satellite lines. On the contrary, the low number of electrons with energies higher than 2~keV results in the lower excitation rates, hence the lower intensities of the allowed lines. As a result, the observed line  
ratio of high $n$ can be obtained. However, the relation between the intensities of the satellite and 
the allowed lines is not so straightforward because electrons with the energy above 2 keV influence also the ionization equilibrium and control the 
\ion{Si}{xiii}/\ion{Si}{xiv} ratio.
The n-distribution alone fulfils the condition of  a strong gradient of the electron distribution function in its high-energy range. 

The electron distribution composed 
of the n-distribution and the Maxwell one does not have 
such a strong gradient in its high-energy part because it has a much higher number of electrons there. Such a composed distribution has the same number of electrons with the energy corresponding the excitation energy of doubly excited state and therefore the same intensities of the satellite lines. 
But, it also has much higher intensities of the allowed lines due to the high number of particles with the energy higher than 2 keV. This must result in 
a much higher \ion{Si}{xiii} to \ion{Si}{xii}d ratio than is observed. The line emission from different locations in the plasma is proportional to their emission measures. The RHESSI spectra indicated that the emission measure of the n-distribution is much higher than for the Maxwell one 
(Tables~\ref{tabx4},~\ref{tabx5}); so the line spectrum of the n-distribution dominates over the Maxwell spectrum, which agrees with the observations \citep{dzi11b}. 
Finally, the electron densities of the power-law distribution, as well as of the electrons producing the type III bursts, are too low
compared to the other two assumed distributions to have any significant effects on the studied RESIK Si line spectrum;
see also the recent results on a combined distribution composed of the n-distribution and the power-law tail \citep{dzi11a}.

In summary, those facts point to different spatial origins for the n-distribution and the thermal
emission.

\section{Conclusions}

In this analysis we have determined that during flares the plasma affected by an injected beam of accelerated electrons can have the non-thermal component in the 2\,--\,6~keV region and that it can be described by the n-distribution. We derived the parameters $n$ and $\tau$ of this n-distribution using the flare spectra from two satellite spectrometers: RESIK and RHESSI. The spectra were
analysed by two distinct diagnostic methods. The soft X-ray lines of \ion{Si}{xiv, xiii, xii}d observed by RESIK provided us with the best possibility 
of analysing the behaviour of parameter $n$. The RHESSI analysis of the n-distribution 
is based on studying its continuous, bremsstrahlung radiation. Although the parameters $n$ and $\tau$ obtained from the RHESSI fits have large uncertainties, their values still overlap with the RESIK results quite well. The RESIK and RHESSI results indicate that departure from the 
Maxwell distribution at 2\,--\,6~keV energies can really occur. The non-thermal n-distribution and the power-law component %were 
are observed in good temporal correlation with the radio bursts detected in the 0.61\,--\,15.40 GHz range by RSTN. The thermal component is still present in RHESSI spectra. RHESSI images show that the emission from both thermal and non-thermal n-distributions comes from nearly the same area. Despite this we argue that in reality each distribution occupies different substructures within the depicted area. Such a conclusion is supported by the character of the observed RESIK line spectra with high fluxes of satellite lines. Finally, our joint analysis of radio and X-ray emission also supports the model of formation of the n-distribution because of the return-current.

\begin{acknowledgements}
This work has been supported by the Scientific Grant Agency VEGA, Slovakia, grant No.~1/0240/11 and Scientific Grant Agency of Czech Republic GA\v{C}R, grants No. 205/09/1705 and P209/10/1680, grant IAA300030701 (GA AS CR), and the research project AV0Z10030501. We are very thankful for the open data policy of RESIK, RHESSI and GOES. BS and JS acknowledge support from the Polish Ministry of Science grant N N203 381736, as well as from the European Commission's Seventh Framework Programme (FP7/2007-2013) under grant agreement No. 218816 (SOTERIA project, http://www.soteria-space.eu).
\end{acknowledgements}

\Online
\begin{appendix}
\section{Complementary figures and tables}
\begin{table*}
\centering
\caption{Time-weighted RESIK spectra for the C5.8 flare on February 22, 2003 and the determined ranges of the parameters $n$ and log~$\tau$.}
   \renewcommand{\arraystretch}{1.2}
\begin{center}
    \begin{tabular}[t]{ccccc}
      \hline \hline
      RESIK spectrum & Time interval & $n$ & log~$\tau$  & GOES phase\\
      &  in UT &  & [K] & \\       \hline
      0  & 09:27:43\,--\,09:28:31 & 4.5\,--\,{\bf 11}\,--\,12 & 7.13\,--\,{\bf 7.25}\,--\,7.26 &  rise\\
      1  & 09:28:31\,--\,09:29:03 & 3.1\,--\,{\bf 5.2}\,--\,12 & 7.07\,--\,{\bf 7.14}\,--\,7.27 &  rise\\
      2  & 09:29:03\,--\,09:30:07 & 2.5\,--\,{\bf 3.0}\,--\,12 & 7.07\,--\,{\bf 7.10}\,--\,7.14 &  maximum\\
      3  & 09:30:07\,--\,09:31:11 & 2.1\,--\,{\bf 2.6}\,--\,3.3 & 7.02\,--\,{\bf 7.05}\,--\,7.08 &  decay\\
      4  & 09:31:11\,--\,09:32:23 & 1.9\,--\,{\bf 2.5}\,--\,3.3 & 6.94\,--\,{\bf 6.99}\,--\,7.03 &  end\\
      5  & 09:32:23\,--\,09:33:51 & 1.5\,--\,{\bf 2.5}\,--\,3.1 & 6.89\,--\,{\bf 6.94}\,--\,6.99 &  post-flare\\
      6  & 09:33:51\,--\,09:35:15 & 1.5\,--\,{\bf 2.4}\,--\,4.0 & 6.87\,--\,{\bf 6.94}\,--\,7.03 &  post-flare\\
      7  & 09:35:15\,--\,09:37:27 & 2.2\,--\,{\bf 3.4}\,--\,9.5 & 6.88\,--\,{\bf 6.97}\,--\,7.15 &  post-flare\\
      \hline \\
      \end{tabular}
\tablefoot{
Parameter values of $n$ and $\tau$ represent the determined values (bold) and their ranges.
}
      \end{center}
    \label{tab1}
\end{table*}

\begin{table*}
\centering
   \caption{Results from RHESSI analysis of the 2003 February 22 flare.}
    \renewcommand{\arraystretch}{1.2}
\begin{center}
    \begin{tabular}[h]{ccccccccccc}
      \hline \hline
       Time interval & RESIK spectrum\tablefootmark{*} & Attenuator & $\chi^{2}_\nu$  & \multicolumn{3}{c}{Thermal parameters}  & \multicolumn{3}{c}{Non-thermal parameters}\\
        in UT & & & & $EM$ & {$kT$} & log~$T$ & $F_\mathrm{T}$ & $\delta$ & $E_\mathrm{C}$\\
         &   &   &   &  [$10^{47}$ cm$^{-3}$] & [keV] & [K] &  [$10^{35}$ s$^{-1}$] &  & [keV]\\
      \hline
      09:28:12\,--\,09:28:20 & 0 & A0 & 0.97 & 0.84 & 1.88 & 7.34 & 0.29 & 2.9 & 13\\
      09:28:28\,--\,09:28:36 & 0, 1 & A1 & 0.90 & 2.9 & 1.87 & 7.34 & 1.1 & 5.4 & 21\\
      09:28:36\,--\,09:28:44 & 1 & A1 & 0.62 & 6.0 & 1.78 & 7.31 & 2.2 & 5.7 & 19\\
      09:28:44\,--\,09:28:52 & 1 & A1 & 1.02 & 10.5 & 1.77 & 7.31 & 3.2 & 6.9 & 19\\
      \hline \\
      \end{tabular}
\tablefoot{
\tablefoottext{*}{See Table~\ref{tab1}.}
}
      \end{center}
    \label{tabx1}
\end{table*}

\begin{table*}
\centering
   \caption{Time-weighted RESIK spectra for the M4.9 flare on January 7, 2003. Notation is the same as in Table~\ref{tab1}.}
    \renewcommand{\arraystretch}{1.2}
\begin{center}
    \begin{tabular}[b]{ccccc}
      \hline \hline
      RESIK spectrum & Time interval & $n$ & log~$\tau$  & GOES phase\\
      &  in UT &  & [K] & \\
      \hline
      0  & 23:13:13\,--\,23:26:33 &  1.0\,--\,{\bf 3.1}\,--\,8.1 & 6.75\,--\,{\bf 6.90}\,--\,7.07 & pre-flare\\
      1  & 23:26:33\,--\,23:28:25 &  2.5\,--\,{\bf 4.2}\,--\,9.6 & 6.87\,--\,{\bf 6.98}\,--\,7.12 & flare beginning \\
      2  & 23:28:25\,--\,23:29:17 &  3.3\,--\,{\bf 6.5}\,--\,12 & 7.04\,--\,{\bf 7.15}\,--\,7.23 & rise\\
      3  & 23:29:17\,--\,23:30:03 &  2.9\,--\,{\bf 3.8}\,--\,6.2 & 7.09\,--\,{\bf 7.13}\,--\,7.19 & rise\\
      4  & 23:30:03\,--\,23:31:03 &  4.6\,--\,{\bf 6.1}\,--\,12 & 7.16\,--\,{\bf 7.20}\,--\,7.29 & rise\\
      5  & 23:31:03\,--\,23:31:51 &  7.7\,--\,{\bf 11}\,--\,12 & 7.25\,--\,{\bf 7.29}\,--\,7.30 & maximum\\
      6  & 23:36:07\,--\,23:36:23 & 11\,--\,{\bf 11}\,--\,12 & 7.22\,--\,{\bf 7.22}\,--\,7.23 & maximum\\
      7  & 23:40:39\,--\,23:41:03 &  4.8\,--\,{\bf 7.3}\,--\,12 & 7.16\,--\,{\bf 7.22}\,--\,7.28 & decay\\
      8  & 23:41:03\,--\,23:42:03 &  3.7\,--\,{\bf 4.6}\,--\,6.0 & 7.14\,--\,{\bf 7.17}\,--\,7.20 & decay\\
      9  & 23:42:03\,--\,23:42:39 &  2.9\,--\,{\bf 3.5}\,--\,4.9 & 7.11\,--\,{\bf 7.13}\,--\,7.17 & decay\\
     10  & 00:21:07\,--\,00:21:55 &  1.6\,--\,{\bf 2.0}\,--\,2.6 & 6.91\,--\,{\bf 6.95}\,--\,6.98 & decay\\
     11  & 00:29:41\,--\,00:30:03 &  1.2\,--\,{\bf 1.9}\,--\,2.8 & 6.82\,--\,{\bf 6.88}\,--\,6.93 & decay\\
     12  & 00:30:03\,--\,00:31:09 &  1.9\,--\,{\bf 2.4}\,--\,2.9 & 6.87\,--\,{\bf 6.91}\,--\,6.94 & decay\\
     13  & 00:31:09\,--\,00:32:23 &  1.0\,--\,{\bf 1.2}\,--\,1.5 & 6.80\,--\,{\bf 6.81}\,--\,6.89 & decay\\
     14  & 00:32:23\,--\,00:33:15 &  1.0\,--\,{\bf 1.9}\,--\,2.5 & 6.78\,--\,{\bf 6.85}\,--\,6.89 & decay\\
     15  & 00:33:15\,--\,00:34:09 &  1.0\,--\,{\bf 2.8}\,--\,11 & 6.75\,--\,{\bf 6.93}\,--\,7.11 & decay\\
      \hline \\
      \end{tabular}
      \end{center}
    \label{tab2}
\end{table*}

\begin{table*}
\centering
   \caption{Results from RHESSI analysis of the 2003 January 7 flare. Notation is the same as in Table~\ref{tabx1}.}
    \renewcommand{\arraystretch}{1.2}
\begin{center}
    \begin{tabular}[t]{ccccccccccc}
      \hline \hline
       Time interval & RESIK Spectrum\tablefootmark{*} & Attenuator & $\chi^{2}_\nu$  & \multicolumn{3}{c}{Thermal parameters}  & \multicolumn{3}{c}{Non-thermal parameters}\\
        in UT & & & & EM & kT & log~T & $F_\mathrm{T}$ & $\delta$ & $E_\mathrm{C}$\\
         &   &   &   &  [$10^{47}$ cm$^{-3}$] & [keV] & [K] &  [$10^{35}$ s$^{-1}$] &  & [keV]\\
      \hline
      23:28:32\,--\,23:28:52 & 2 & A0 & 1.07 & 0.64 & 1.96 & 7.36 &  3.4 & 7.0 & 16\\
      23:29:16\,--\,23:29:36 & 3 & A1 & 1.05 & 4.7 & 1.73 & 7.30 &  2.9 & 7.0 & 19\\
      23:29:36\,--\,23:30:00 & 3 & A1 & 0.58 & 5.6 & 2.06 & 7.38 &  2.7 & 6.9 & 24\\
      23:30:00\,--\,23:30:28 & 4 & A1 & 0.89 & 19. & 1.81 & 7.32 & 10.5 & 6.3 & 19\\
      23:30:28\,--\,23:31:00 & 4 & A1 & 0.94 & 50. & 1.69 & 7.29 & 26. & 7.3 & 18\\
      23:31:00\,--\,23:31:44 & 5 & A1 & 1.01 & 67. & 1.89 & 7.34 & 27. & 7.1 & 20\\
      23:32:00\,--\,23:32:32 &\,--\,& A3 & 0.57 & 45. & 2.38 & 7.44 &  2.9 & 6.2 & 26\\
      23:32:32\,--\,23:33:00 &\,--\,& A3 & 0.84 & 19. & 1.90 & 7.34 & 18. & 8.6 & 22\\
      \hline \\
      \end{tabular}
\tablefoot{
\tablefoottext{*}{See Table~\ref{tab2}.}
}
      \end{center}
    \label{tabx2}
\end{table*}

\begin{table*}
\centering
    \caption{Time-weighted RESIK spectra for the M4.0 flare on October, 4, 2002. Notation is the same as in Table~\ref{tab1}.}
    \renewcommand{\arraystretch}{1.2}
\begin{center}
    \begin{tabular}[h]{ccccc}
      \hline \hline
      RESIK spectrum & Time interval  & $n$ & log~$\tau$  & GOES phase\\
      &  in UT &  & [K] & \\       \hline
      0  & 05:35:19\,--\,05:36:03  &   3.5\,--\,{\bf 5.5}\,--\,12 & 7.07\,--\,{\bf 7.14}\,--\,7.26 & rise\\
      1  & 05:36:03\,--\,05:36:35  &   2.1\,--\,{\bf 2.7}\,--\, 3.8 & 6.97\,--\,{\bf 7.0}1\,--\,7.06 & rise\\
      2  & 05:36:35\,--\,05:37:01  & 11\,--\,{\bf 11}\,--\,12 & 7.21\,--\,{\bf 7.21}\,--\,7.22 & rise\\
      3  & 05:37:01\,--\,05:37:31  &  3.1\,--\,{\bf 7.6}\,--\,12 & 7.09\,--\,{\bf 7.22}\,--\,7.28 & rise\\
      4  & 05:37:31\,--\,05:38:01  &  2.4\,--\,{\bf 4.4}\,--\,12 & 7.05\,--\,{\bf 7.14}\,--\,7.29 & maximum\\
      5  & 05:38:01\,--\,05:38:31  & 11\,--\,{\bf 11}\,--\,12 & 7.24\,--\,{\bf 7.24}\,--\,7.25 & maximum\\
      6  & 05:38:31\,--\,05:39:01  & 11\,--\,{\bf 11}\,--\,12 & 7.23\,--\,{\bf 7.23}\,--\,7.24 & decay\\
      7  & 05:39:01\,--\,05:39:31  &  2.6\,--\,{\bf 4.2}\,--\,12 & 7.09\,--\,{\bf 7.16}\,--\,7.29 & decay\\
      8  & 05:39:31\,--\,05:40:01  &  4.2\,--\,{\bf 9.3}\,--\,12 & 7.12\,--\,{\bf 7.23}\,--\,7.28 & decay\\
      9  & 05:40:01\,--\,05:41:00  &  2.5\,--\,{\bf 3.1}\,--\,4.0 & 7.07\,--\,{\bf 7.10}\,--\,7.14 & decay\\
     10  & 05:41:00\,--\,05:42:00  &  4.7\,--\,{\bf 6.5}\,--\,12 & 7.16\,--\,{\bf 7.20}\,--\,7.29 & end\\
     11  & 05:42:00\,--\,05:43:00  &  3.3\,--\,{\bf 4.1}\,--\,5.9 & 7.09\,--\,{\bf 7.12}\,--\,7.17 & post-flare\\
     12  & 05:43:00\,--\,05:44:00  &  2.8\,--\,{\bf 3.3}\,--\,4.1 & 7.08\,--\,{\bf 7.10}\,--\,7.13 & post-flare\\
     13  & 05:44:00\,--\,05:45:00  &  1.9\,--\,{\bf 2.1}\,--\,2.8 & 7.06\,--\,{\bf 7.08}\,--\,7.11 & post-flare\\
     14  & 05:45:00\,--\,05:46:00  &  2.7\,--\,{\bf 3.1}\,--\,3.8 & 7.06\,--\,{\bf 7.08}\,--\,7.11 & post-flare\\
     15  & 05:46:00\,--\,05:47:02  &  2.6\,--\,{\bf 3.0}\,--\,3.6 & 7.03\,--\,{\bf 7.05}\,--\,7.08 & post-flare\\
     16  & 05:47:02\,--\,05:48:02  &  1.8\,--\,{\bf 2.2}\,--\,2.6 & 6.98\,--\,{\bf 7.00}\,--\,7.02 & post-flare\\
     17  & 05:48:02\,--\,05:49:02  &  2.6\,--\,{\bf 2.9}\,--\,3.5 & 7.00\,--\,{\bf 7.03}\,--\,7.05 & post-flare\\
     18  & 05:49:02\,--\,05:50:02  &  3.0\,--\,{\bf 3.5}\,--\,5.0 & 7.03\,--\,{\bf 7.06}\,--\,7.11 & post-flare\\
     19  & 05:50:02\,--\,05:51:06  &  1.2\,--\,{\bf 1.6}\,--\,2.0 & 6.93\,--\,{\bf 6.96}\,--\,6.98 & post-flare\\
     20  & 05:51:06\,--\,05:52:02  &  1.7\,--\,{\bf 2.1}\,--\,2.5 & 6.92\,--\,{\bf 6.95}\,--\,6.98 & post-flare\\
     21  & 05:52:02\,--\,05:53:06  &  1.9\,--\,{\bf 2.3}\,--\,2.7 & 6.92\,--\,{\bf 6.95}\,--\,6.98 & post-flare\\
     22  & 05:53:06\,--\,05:54:06  &  2.6\,--\,{\bf 3.1}\,--\,4.4 & 6.99\,--\,{\bf 7.02}\,--\,7.08 & post-flare\\
      \hline \\
      \end{tabular}
      \end{center}
    \label{tab3}
\end{table*}

\begin{table*}
\centering
   \caption{Results from RHESSI analysis of the 2002 October 4 flare. Notation is the same as in Table~\ref{tabx1}.}
    \renewcommand{\arraystretch}{1.2}
\begin{center}
    \begin{tabular}[b!]{ccccccccccc}
      \hline \hline
       Time interval & RESIK spectrum\tablefootmark{*} & Attenuator & $\chi^{2}_\nu$  & \multicolumn{3}{c}{Thermal parameters}  & \multicolumn{3}{c}{Non-thermal parameters}\\
        in UT & & & & EM & kT & log~T & $F_\mathrm{T}$ & $\delta$ & $E_\mathrm{C}$\\
         &   &   &   &  [$10^{47}$ cm$^{-3}$] & [keV] & [K] &  [$10^{35}$ s$^{-1}$] &  & [keV]\\ 
      \hline
      05:35:28\,--\,05:36:00 & 0 & A1 & 1.18 & 9.6 &  1.84 & 7.33 & 586. & 5.8 & 8\\
      05:36:00\,--\,05:36:16 & 1 & A1 & 1.13 & 51. &  1.89 & 7.34 &  27.1 & 7.6 & 19\\
      05:36:32\,--\,05:37:00 & 2 & A3 & 0.88 & 63. &  1.92 & 7.35 &  10.3 & 7.9 & 22\\
      05:37:00\,--\,05:37:32 & 3 & A3 & 0.87 & 105. &  1.86 & 7.33 &  10.5 & 10.4 & 23\\
      \hline \\
      \end{tabular}
\tablefoot{
\tablefoottext{*}{See Table~\ref{tab3}.}
}
      \end{center}
    \label{tabx3}
\end{table*}

\begin{table*}
\centering
\caption{
Details of the RHESSI spectrum fitting (2003 February 22, 09:28:12\,--\,09:28:20 UT) and comparison with the corresponding RESIK results
(09:27:43\,--\,09:28:31 UT).}
    \renewcommand{\arraystretch}{1.2}
\begin{center}
    \begin{tabular}[t]{ccccc}
      \hline \hline
       Model & Parameter & \multicolumn{3}{c}{Parameter values}\\
      \hline
       \,--\, & Energy range [keV] & 6\,--\,100  & 4\,--\,100 & 4\,--\,100\\
       \,--\,    & $\chi^{2}_\nu$ & 0.97 & 7.33 & 1.12\\
      \hline
      vth & $EM$ [$10^{47}$ cm$^{-3}$] & 0.84 & 2.2 & 0.70\,--\,{\bf 0.85}\,--\,0.90\\
      vth & $kT$ [keV] & 1.88 & 1.45 & 1.83\,--\,{\bf 1.86}\,--\,1.91\\
      \hline
      photon\_thick & $F_\mathrm{T}$ [$10^{35}$ electrons.s$^{-1}$] & 0.29 & 1.2 & 0.09\tablefootmark{a}\,--\,{\bf 0.47}\\
      photon\_thick & $\delta$ & 2.9 & 3.2  & 2.8\,--\,{\bf 2.9}\,--\,3.0\\
      photon\_thick & $E_\mathrm{C}$ [keV]  & 13 & 8.9 & {\bf 9.8}\,--\,32\tablefootmark{b} \\
      \hline
      thin\_ndist & $EM_n$ [$10^{47}$ cm$^{-3}$] &\,--\,&\,--\,& 5.0\tablefootmark{a}\,--\,{\bf 5.1}\\
      thin\_ndist & $n$ &\,--\,&\,--\,& 2\,--\,9 ({\bf 11}\tablefootmark{c})\\
      thin\_ndist & $k\tau$ [keV] &\,--\,&\,--\,& 1.0\,--\,1.5 ({\bf 1.63}\tablefootmark{c} )\\
      \hline
      RESIK & $n$ &\,--\,&\,--\,& 4.5\,--\,{\bf 11}\,--\,12\\
      RESIK & $k\tau$ [keV] &\,--\,&\,--\,& 1.16\,--\,{\bf 1.53}\,--\,1.57\\
      \hline \\
      \end{tabular}
\tablefoot{
Columns three to five list parameter values of three types of RHESSI fits (original, i.e. vth + photon-thick, 
original within the enlarged energy range,
and the one combined with the n-distribution), see Sect.~\ref{sect:rhessi_ndistr} and left panels of 
Figs.~\ref{fig_rhessi_o}, \ref{fig_rhessi_w}, and \ref{fig_rhessi_n}.
Last column shows the fitted parameters (bold), together with their limits.
\tablefoottext{a}{lower limit only}
\tablefoottext{b}{upper limit only}
\tablefoottext{c}{values outside estimated 1$\sigma$ range}
}
    \end{center}
    \label{tabx4}
\end{table*}

\begin{table*}
\centering
\caption{
Details of the RHESSI spectrum fitting (2003 January 7, 23:28:32\,--\,23:28:52 UT) and comparison with the corresponding RESIK results
(23:28:25\,--\,23:29:17 UT). Notation is the same as in Table~\ref{tabx4}. }
    \renewcommand{\arraystretch}{1.2}
\begin{center}
    \begin{tabular}[h]{ccccc}
      \hline \hline
      Model & Parameter & \multicolumn{3}{c}{Parameter values}\\
      \hline
     \,--\,& Energy range [keV] & 6\,--\,40  & 4\,--\,40 & 4\,--\,40\\
      \,--\,& $\chi^{2}_\nu$ & 1.07 & 5.38 & 1.31\\
      \hline
      vth & $EM$ [$10^{47}$ cm$^{-3}$] & 0.64 & 4.1 & 0.3\,--\,{\bf 0.35}\,--\,0.41\\
      vth & $kT$ [keV] & 1.96 & 1.18 & 2.3\,--\,{\bf 2.44}\,--\,2.7\\
      \hline
      photon\_thick & $F_\mathrm{T}$ [$10^{35}$ electrons.s$^{-1}$] & 3.4 & 9.6 & 1.6\,--\,{\bf 2.4}\,--\,3.1\\
      photon\_thick & $\delta$ & 7.0 & 7.2  & 5.8\,--\,{\bf 6.5}\,--\,6.7\\
      photon\_thick & $E_\mathrm{C}$ [keV]  & 16 & 14 & 15\,--\,{\bf 15.5}\,--\,16\\
      \hline
      thin\_ndist & $EM_\mathrm{n}$ [$10^{47}$ cm$^{-3}$] &\,--\,&\,--\,& 0.5\tablefootmark{a}\,--\,{\bf 4.5} \\
      thin\_ndist & $n$ &\,--\,&\,--\,& 5.0\tablefootmark{a}\,--\,{\bf 6.5} \\
      thin\_ndist & $k\tau$ [keV] &\,--\,&\,--\,& 1.5\,--\,{\bf 1.72}\,--\,2.6\\
      \hline
      RESIK & $n$ &\,--\,&\,--\,& 3.3\,--\,{\bf 6.5}\,--\,12\\
      RESIK & $k\tau$ [keV] &\,--\,&\,--\,& 0.94\,--\,{\bf 1.22}\,--\,1.46\\
      \hline \\
      \end{tabular}
\tablefoot{
See also Sect.~\ref{sect:rhessi_ndistr} and right panels of 
Figs.~\ref{fig_rhessi_o}, \ref{fig_rhessi_w}, and \ref{fig_rhessi_n}.\\
\tablefoottext{a}{lower limit only}
}
      \end{center}
    \label{tabx5}
\end{table*}

\begin{figure*}[t]
\centering {\includegraphics[angle=90,width=8.5cm]{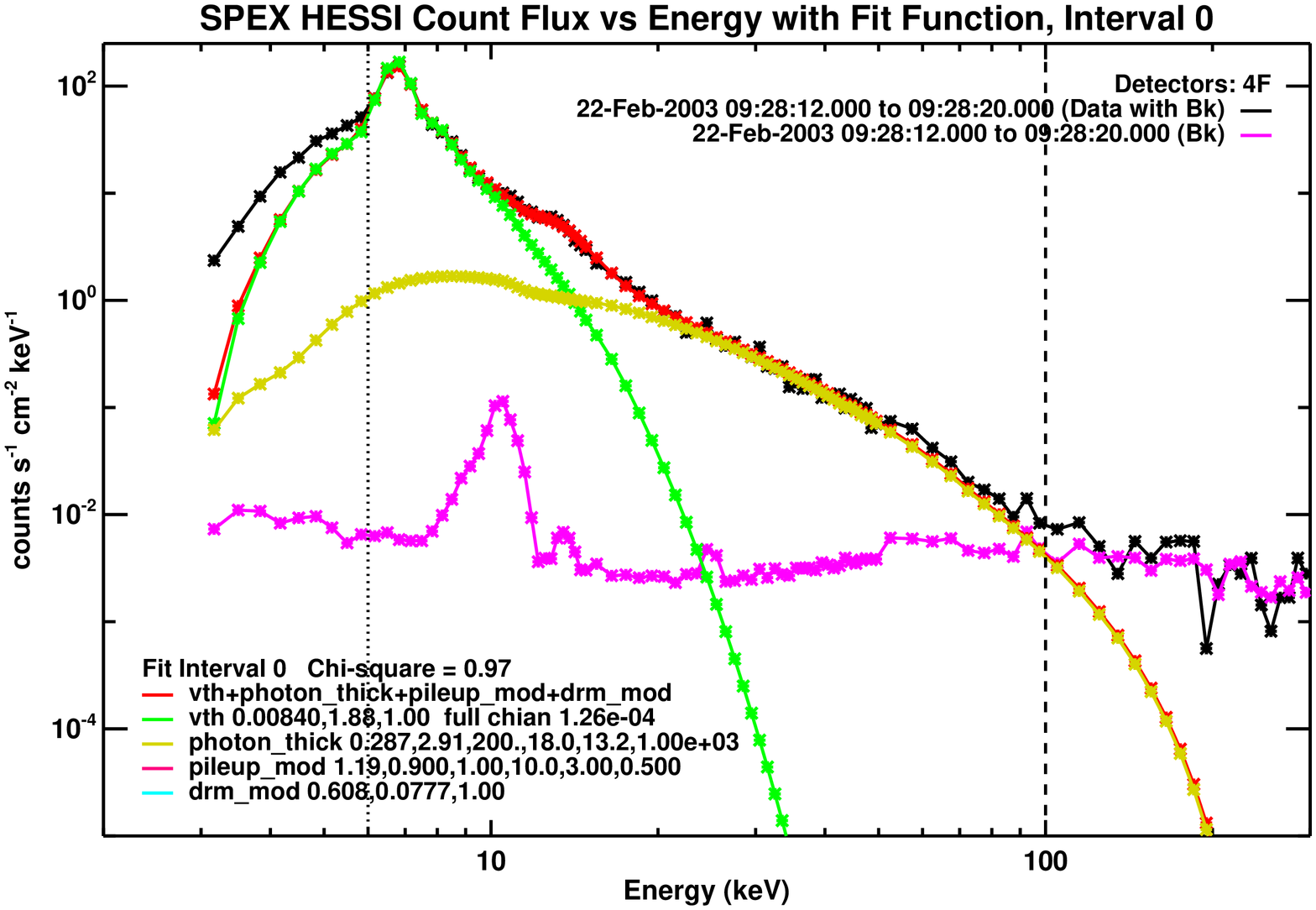}\includegraphics[angle=90,width=8.5cm]{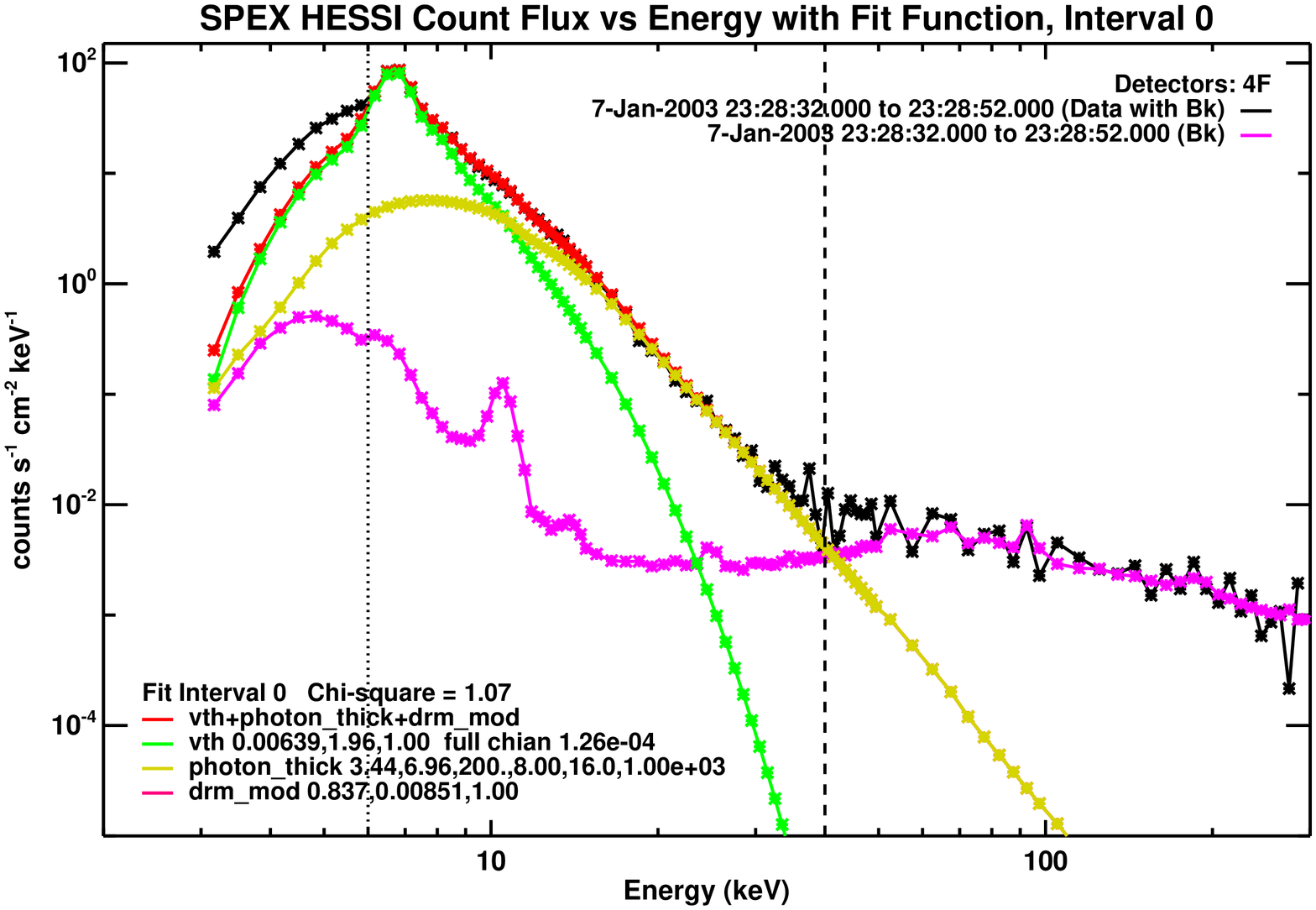}}
{\includegraphics[angle=90,width=8.5cm]{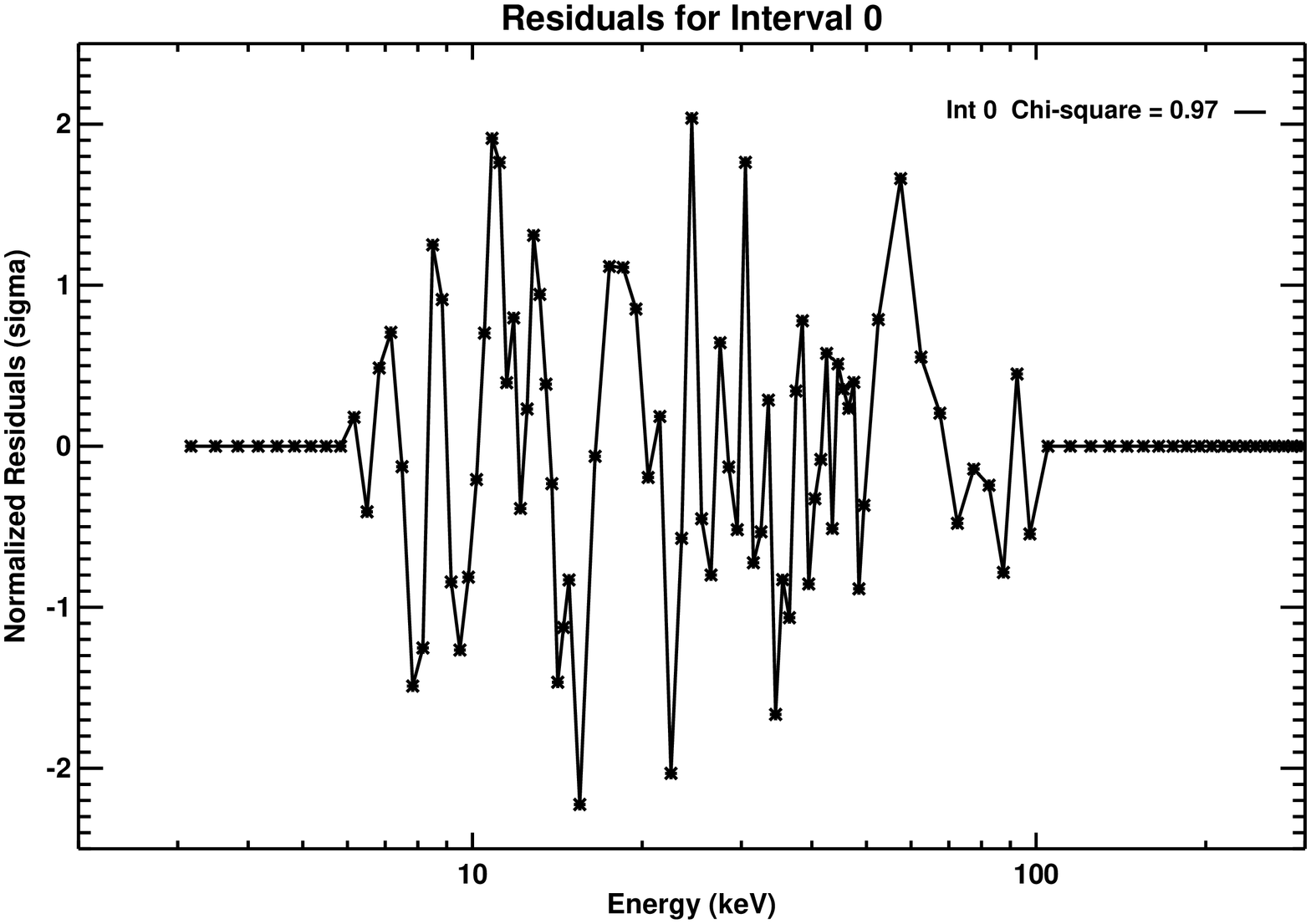}\includegraphics[angle=90,width=8.5cm]{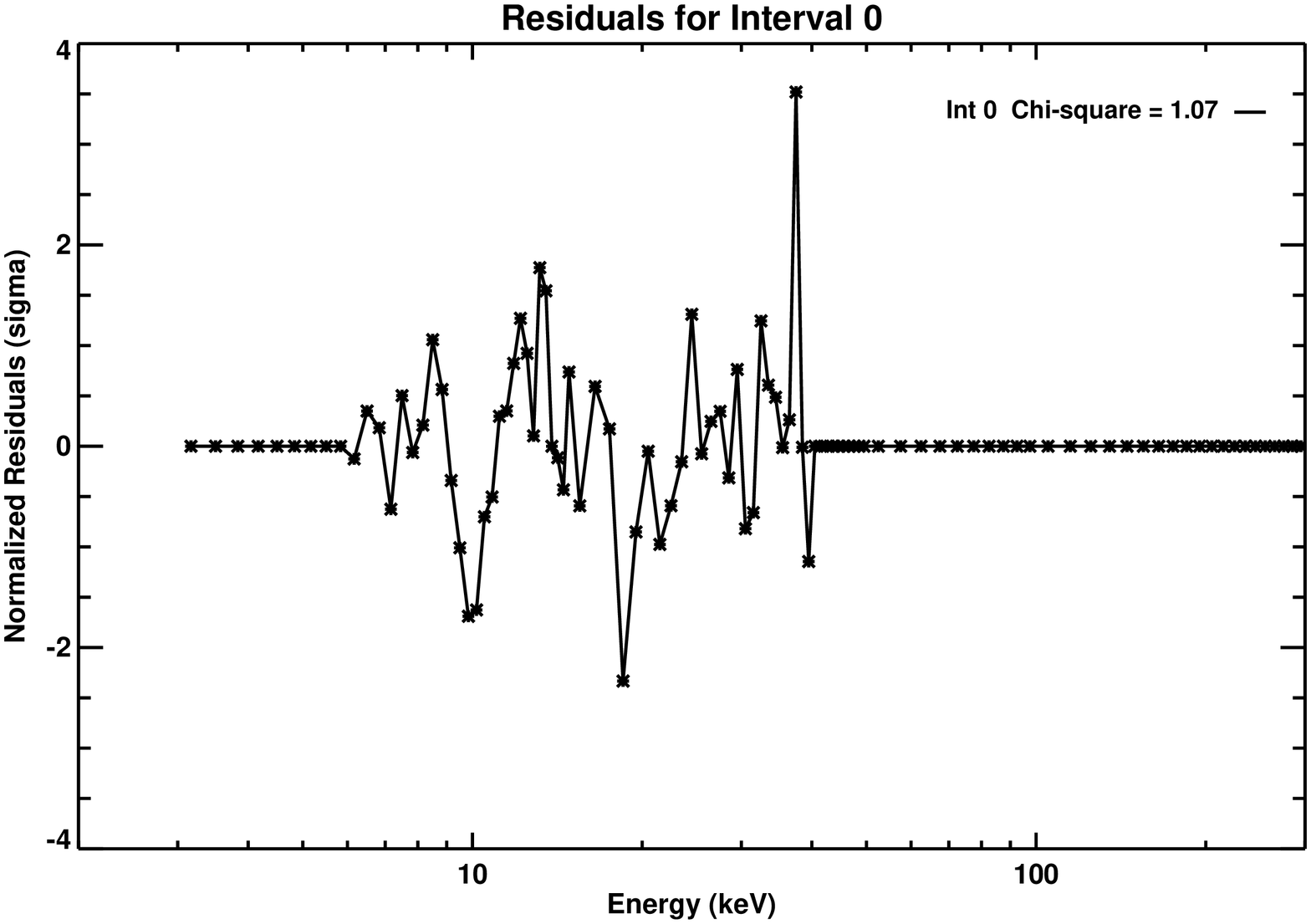}}
\caption{RHESSI fits (from 6 keV) of the single power law (thick-target model) and 
the thermal component together with residuals: the 2003 February 22 flare (left) and the 2003 January 7  flare (right).}
\label{fig_rhessi_o}
\end{figure*}

\begin{figure*}[b]
\centering {\includegraphics[angle=90,width=8.5cm]{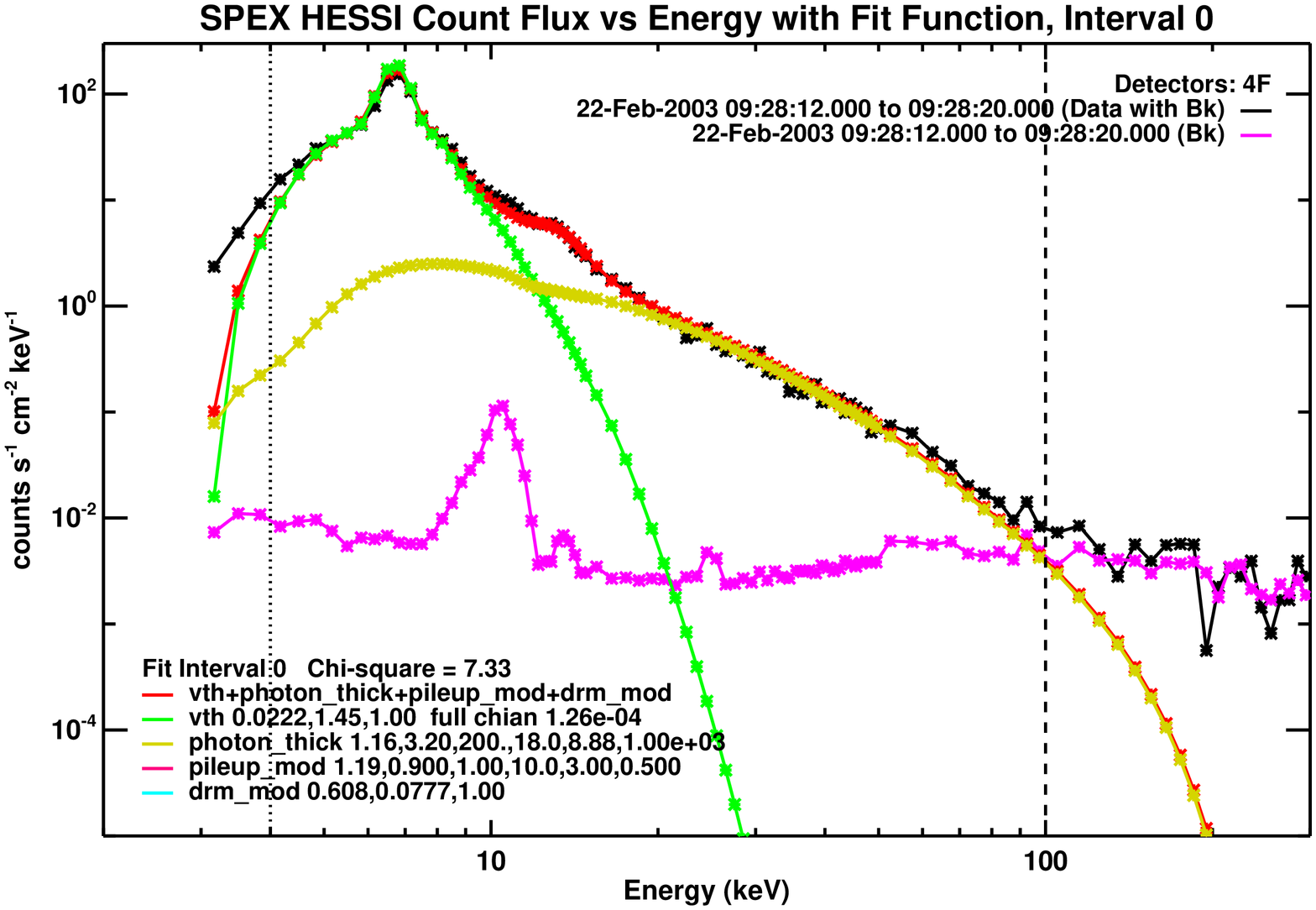}\includegraphics[angle=90,width=8.5cm]{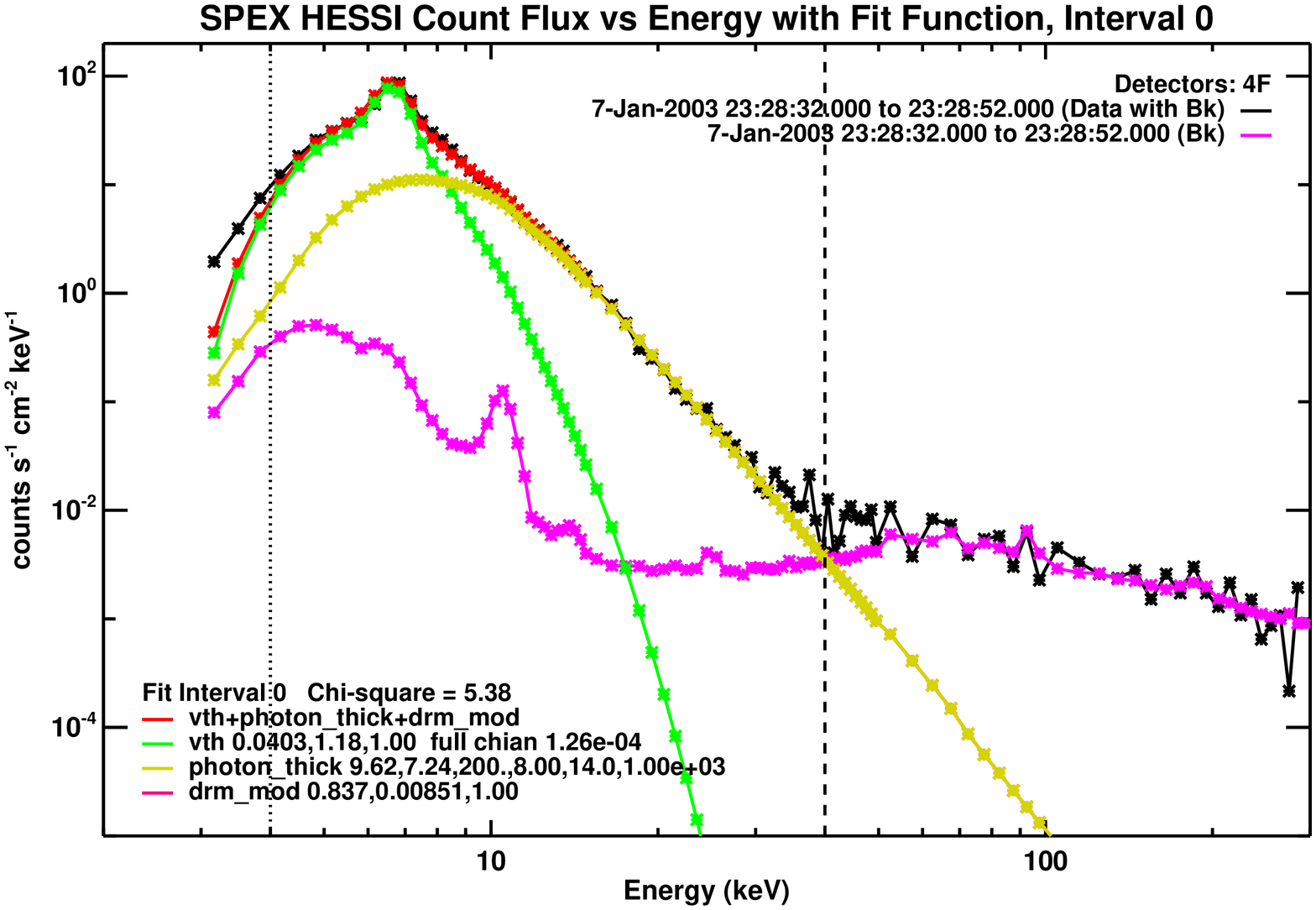}}
{\includegraphics[angle=90,width=8.5cm]{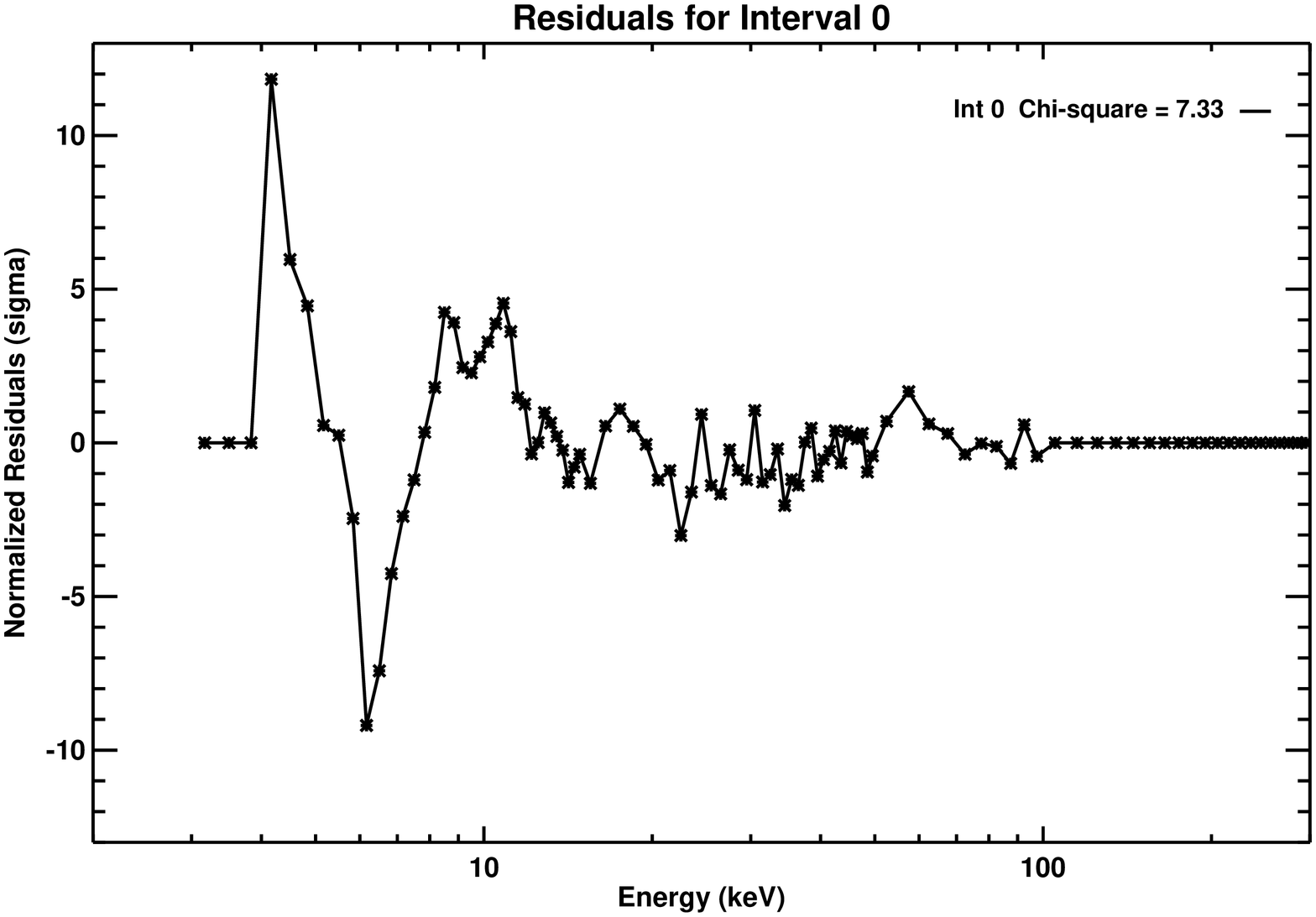}\includegraphics[angle=90,width=8.5cm]{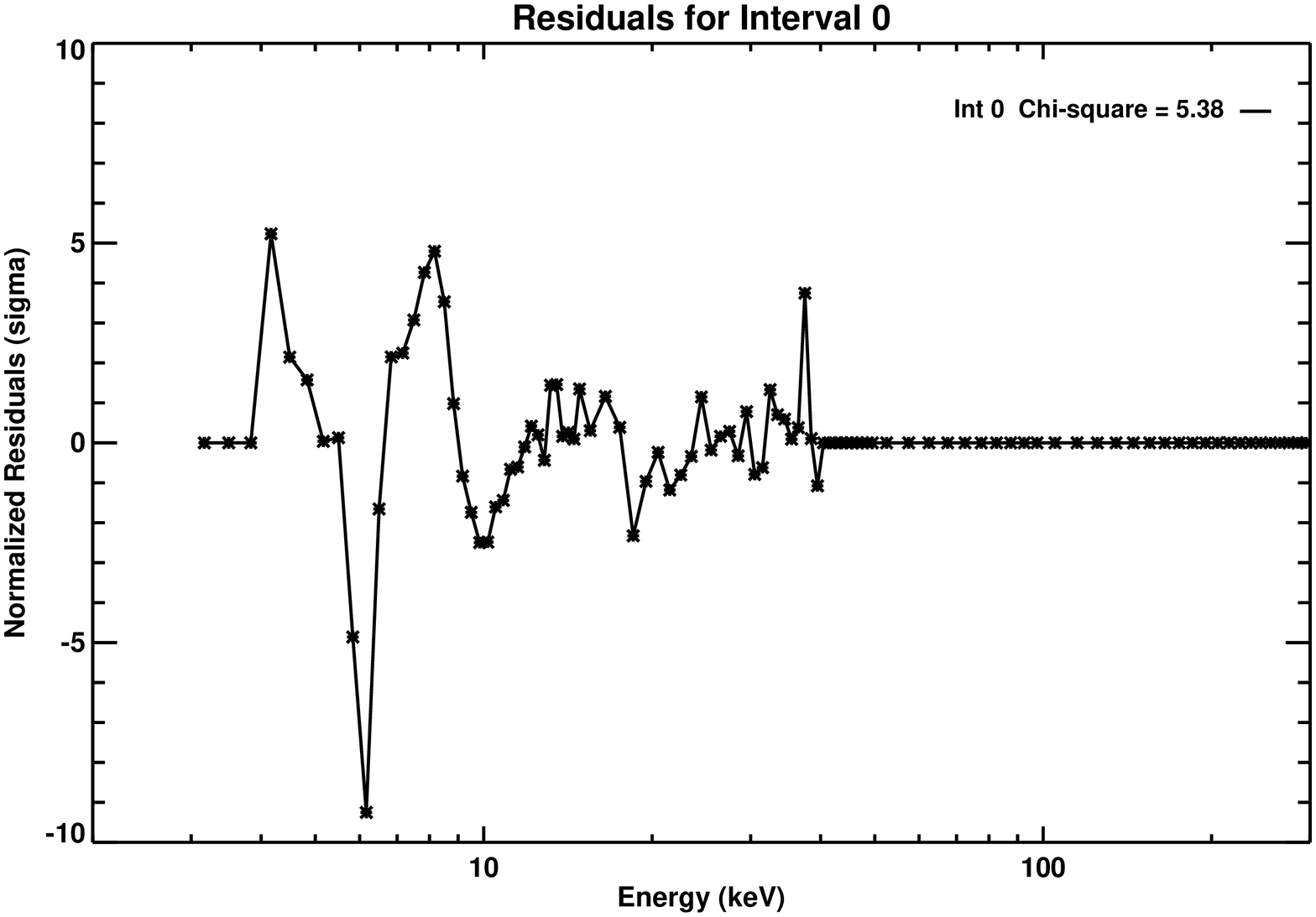}}
\caption{Widened RHESSI fits (from 4 keV). Notation is the same is in Fig.~\ref{fig_rhessi_o}.}
\label{fig_rhessi_w}
\end{figure*}

\end{appendix}
\end{document}